\documentclass[11pt,a4paper]{article}
\usepackage{graphicx}      
\usepackage{booktabs}       
\usepackage{natbib}        
\usepackage{hyperref}      
\usepackage[labelfont=bf]{caption}
\usepackage[utf8]{inputenc}
\usepackage[T1]{fontenc}
\usepackage{lmodern}
\usepackage{amsmath,amssymb,amsfonts}
\usepackage{multirow}
\usepackage{tabularx}
\usepackage{threeparttable} 
\usepackage{xcolor}
\usepackage{microtype}
\usepackage{subcaption}
\usepackage{enumitem}
\usepackage{titlesec}
\usepackage{appendix}
\usepackage{float}
\usepackage{array}
\usepackage{url}

\newcolumntype{L}{>{\raggedright\arraybackslash}X}

\usepackage[margin=1in]{geometry}
\usepackage{setspace}
\hypersetup{
    colorlinks=true,
    linkcolor=blue!80!black,
    citecolor=blue!80!black,
    urlcolor=blue!70!black,
    pdftitle={AI in Science: Early Insights},
    pdfauthor={Mihai Codreanu, Alex Imas, Juan Mateos-Garcia et al.}
}

\title{
    \vspace{-2.2em}
    \textbf{
    \huge AI in Science: Early Insights\thanks{This paper introduces a new Google and Google DeepMind research effort, tracking AI use and (economic) impact in science. This first snapshot is the result of a collaboration between Google ATLAS, Google DeepMind and MIT FutureTech. We are thankful for help, feedback and/or support from Lizzie Dorfman, Alex Goldin, Conor Griffin, Alon Halevy, Demis Hassabis, Steph Hughes-Fitt, Andrew Kim, Pushmeet Kohli, Shane Legg, Anu Madgavkar, Mike Pisa, Priya Ramu, Philippa Rock, Vivek Sampathkumar, Alison Snyder, Leanne Trujillo.}
    }
\vspace{-0.5em} }

\author{
    \fontsize{11.5pt}{13.5pt}\selectfont 
    \begin{tabular}{c @{\hspace{1.8em}} c @{\hspace{1.8em}} c @{\hspace{1.8em}} c}
        Mihai Codreanu$^{1,}$\thanks{Mihai Codreanu (\href{mailto:mihaicod@google.com}{mihaicod@google.com}), Alex Imas (\href{mailto:imasa@google.com}{imasa@google.com}), and Juan Mateos-Garcia (\href{mailto:jmateosgarcia@google.com}{jmateosgarcia@google.com}) co-led this research, have contributed equally and are all corresponding authors. Order was decided purely alphabetically.} & Alex Imas$^{2,3,\dagger}$ & Juan Mateos-Garcia$^{2,\dagger}$ & Joseph Emmens$^{4,5}$ \\[0.15em]
        Evalyne Muiruri$^2$ & Arthur Turrell$^1$ & Julian Jacobs$^{2,6}$ & Atoosa Kasirzadeh$^{2,7}$ \\[0.15em]
        Ana Tri\v{s}ovi\'{c}$^5$ & Yiyuan Chen$^1$ & Tanya Rodchenko$^1$ & Catherine Pollard$^2$ \\[0.15em]
        Scott Strand$^1$ & Daniel Rock$^{1,8}$ & Zanna Iscenko$^1$ & Fabien Curto Millet$^1$ \\[0.15em]
        \multicolumn{4}{c}{Neil Thompson$^5$ \hspace{1.8em} James Manyika$^1$}
    \end{tabular} \\[0.6em]
    \begin{tabular}{c}
        \fontsize{9pt}{11.5pt}\selectfont $^1$Google \quad $^2$Google DeepMind \quad $^3$University of Chicago \quad $^4$CUNEF Universidad \\[0.1em]
        \fontsize{9pt}{11.5pt}\selectfont $^5$MIT FutureTech \quad $^6$University of Oxford \quad $^7$Carnegie Mellon University \quad $^8$University of Pennsylvania
    \end{tabular}
}

\vspace{-2em} 
\date{\normalsize September 2026}

\begin{document}

\maketitle

\vspace{-3em} 

\begin{abstract}
\noindent Scientific progress is a key driver of economic growth and prosperity. There is great excitement - but also concerns - about the impacts of AI on science, but so far little data. We provide early insights on this from three data sources: a sample of 15 million Gemini interactions, an inventory of over 2,600 specialized AI models across disciplines, and a survey of over 600 scientists. We map these data to a new taxonomy of scientific tasks to study how scientists are using AI. Four main findings emerge. First, we find broad adoption and coverage: scientists use AI more than most other occupations. Specialized AI models have broad disciplinary coverage and are highly cited. Nearly half of the scientists surveyed report using some form of AI every day. Second, we document evidence that LLMs (proxied through Gemini usage) and specialized models act as complements---LLMs are used for general analysis, coding, and manuscript preparation, while specialized models provide domain-specific predictions, data generation and classification. Third, scientists report large productivity gains from using AI: a saving of nearly 7 hours per week, time which is primarily re-invested in more research. Finally, we show that AI is already changing the scientific process. As some stages of scientific research become easier, bottlenecks shift downstream. Scientists report an increased backlog of untested hypotheses and substantial demand for output verification. Our findings suggest that AI holds significant potential to increase scientific productivity. However, as with other sectors, its ultimate impact will be governed by complex task interdependencies and investment into the elimination of emerging bottlenecks.
\end{abstract}

\noindent\textbf{Keywords:} artificial intelligence, economics of science, productivity, economic growth.\\[0.4em]
\textbf{JEL Classification Codes:} O33, O31, O32, J24
\newpage

\section{Introduction}\label{sec:introduction}

Scientific discovery is one of the primary drivers of long-run progress and economic growth \citep{romer1990endogenous, mokyr1992lever, aghion1992model, jones1995rd, akcigit2020back}. Yet concerns of decreased research productivity or ``ideas getting harder to find'' have been raised in recent years \citep{Bloom2020, Park2023}. The rapid development of AI has fueled optimism that the technology could reverse this trend and speed up scientific progress \citep{wang2023scientific, amodei2024machines, curtomillet2025will, hassabis2026ai, agrawal2026ai}. Specialized AI models, such as DeepMind's AlphaFold 2 \citep{jumper2021highly, hassabis2022alphafold} have already been recognized as Nobel-caliber breakthroughs. Meanwhile, general-purpose LLMs are rapidly diffusing across day-to-day research workflows \citep{VanNoorden2023, wiley2024explanations}. 

If AI acts as an ``IMI'', an Invention of a Method of Invention \citep{aghion2017artificial, cockburn2018impact, crafts2021artificial, cunningham2026economics}, it would create a durable acceleration to productivity growth over time.\footnote{Even short of being an ``IMI'', an increase in the rate of growth could happen even if it just allows scientists to harness more computational capital in their research \citep{besiroglu2023economicimpactsaiaugmentedrd}. At the other end of the spectrum, there is increasing discussion of recursive self-improvement or ``RSI'', a stage at which AI models will not only help speed up scientists' workflow but will be able to design, code, or even train their own successors \citep[e.g.,][]{good1965speculations, huang2023large, lu2024aiscientist}.} Much is riding on AI's potential for economic growth and progress, as increasing government debt and aging populations threaten to disrupt the balanced growth path of roughly 2 percent annual expansion per capita over the last 150 years in the US \citep{jones2026ai}; AI's role in science may be key for overcoming these headwinds and accelerating growth.

Yet the optimism around AI's potential to increase scientific progress is not universal. Some worry that a deluge of AI-generated content, hallucinations, and ``illusions of understanding'' will strain peer review and validation \citep{birhane2023science,messeri2024artificial}. Others argue that models optimized on the existing literature will steer researchers toward incremental, well-known paradigms rather than high-risk, high-reward discoveries \citep{duede2025ai}, or that steep compute and capital requirements will widen the gap between well-funded laboratories and under-resourced institutions \citep{thompson2022computationallimitsdeeplearning,Ahmed2023, besiroglu2024computedividemachinelearning}. Some also worry that AI might accelerate the production of low-quality research that merely appears scientific \citep{luo2025more, gyevnar2026scientific}. But these debates have thus far evolved with little empirical evidence.\footnote{While an emerging literature has begun characterizing AI's role as an emerging ``general method of invention'' \citep{bianchini2022artificial, bianchini2025drivers}, documenting some of the rapid cross-disciplinary diffusion and citation premia of AI research \citep{gao2024quantifying}, examining the implications of automating research \citep{duede2024after, duede2025ai} and emerging impacts of significant developments such as AlphaFold 2 \citep{hill2026artificial, gans2026how, Qian2026, cavalli2026how}, systematic empirical evidence on real-time, day-to-day adoption remains scarce.} 

Prior academic research about AI adoption and its impact on science has primarily relied on open scientometric records such as publication and citation trends, patent filings, and traces of LLM-assisted writing \citep{kusumegi2025scientific, renault2026commentscientificproductionera, trisovic2025, duede2024oil}. These are finalized outputs that arrive with a lag and track adoption imperfectly. Frontier AI labs hold real-time telemetry on how researchers actually use models, but the studies built on such logs have so far focused on aggregate labor market patterns \citep{chatterji2025chatgpt, appel2025anthropic, handa2025economictasksperformedai, iscenko2026atlas}, with relatively little focus on science.\footnote{Recent work on potential impacts especially has focused on generative AI as a category including chatbots and early examples of AI agents \citep{eloundou2024gpts, massenkoff2026, chatterji2025chatgpt}.} 

This paper inaugurates a new research program on the impact of AI on science by combining insights from three new data sources: a sample of 15 million Gemini interactions, bibliometrics data for more than 2,600 specialized AI models across academic disciplines (from protein structure prediction to materials discovery and weather forecasting), and a survey of more than 600 scientists. We map all three data sources to a new taxonomy of scientific tasks developed by MIT FutureTech, which gives task-level granularity across OpenAlex scientific subfields \citep{emmens2026taxonomy}. 

The analysis reveals four findings. First, scientists lead other occupations in the use of AI; scientific occupations are overrepresented in Gemini usage relative to their share of employment. Nearly half of surveyed scientists report using AI every day as part of their workflow. But looking at LLM use alone would miss much of the picture. Surveyed scientists use specialized models about as much as coding assistants. These models are also heavily cited in scholarly work. AI adoption in science is also geographically concentrated, with usage scaling with a country's scientific workforce. Second, LLMs and specialized models are economic complements with a clear division of labor between them: the former are primarily used broadly across tasks including coding and writing, while the latter are used for more specialized generation, prediction, and simulation, with limited overlap in tasks between the two model categories. Third, scientists report substantial productivity gains from AI, with average time saving of just below 7 hours per week. The time is mostly put back into research. Researchers also report greater cross-field and interdisciplinary insights. Fourth, while some tasks---for example, quantitative, computational work---may be accelerated, the bottlenecks to scientific output shift downstream into physical experimentation and validation. Scientists report a backlog of untested hypotheses, substantial time spent verifying AI outputs, and a tilt toward safer questions.

To give more detail on the methodology, our first data source comes from the Google ATLAS 1.0 project \citep{iscenko2026atlas} which consists of about 15 million anonymized interactions across the conversational surfaces (Gemini App and AI Mode), and Gemini API. Scientific work is isolated through a three-stage pipeline. First, a classifier removes non-work and educational interactions; then, a filter restricts the sample to the detailed occupations where research is concentrated; finally, as these occupations often contain non-scientists, a new science classifier is built on the \citet{oecd2015frascati} and \citet{ukri2025definition} using the content of interactions to classify if they are likely part of a scientist workflow. This leaves us with 360,000 science interactions. 

But how does one map LLM interactions to specific tasks? Standard taxonomies exist for occupational tasks (O*NET) and time use (ATUS). But when applied to scientific research, these tools are insufficient. Generic categories like ``analyzing data'' or ``writing reports'' flatten the domain-specific activities that define research, from hypothesis formulation to molecular construct design or assay optimization. Measuring AI's impact on science requires a common metric that separately captures what researchers do across disciplines and which of those tasks AI tools help, enable, or automate. To accomplish this, we used our custom hierarchical discovery and classification engine - Observation Clustering and Taxonomy Organisation (OCTO, \cite{iscenko2026atlas}) - to map scientific LLM interactions to two science-based taxonomies: the OpenAlex scientific disciplines and the three levels of the new MIT FutureTech task taxonomy. 

The second data source comes from a newly-compiled inventory of over 2,600 specialized models published since 2012. These include models from many disciplines including biology (e.g., AlphaFold \citep{hassabis2022alphafold} predicting protein folding), material science (e.g., GNoME \citep{Merchant2023} predicting stable crystal structures), chemistry (e.g., MatterGen \citep{Zeni2025}, a generative diffusion model for inorganic compound design) and more. We assembled the inventory by combining a bottom-up deep agentic search over web sources, publications, and code repositories with Epoch's AI model database \citep{epochai2026models}, then enriched the output using metadata from OpenAlex. OCTO was then used to extract the research tasks that each model enables (e.g., predicting protein structures), which were then mapped to the respective tasks in the FutureTech and OpenAlex taxonomies, as done with the log data. This exercise puts the LLM and specialized model usage on a common metric, which allows us to empirically study the division of labor between these categories of models. We have tracked 460,000 unique citations to these models to trace diffusion and knowledge flows across scientific fields.

Our final data source is an original survey of 637 active researchers in the US and UK carried out in July-August 2026. The sample included Principal Investigators (PIs) in academia, industry researchers, and early career scientists across four major scientific domains. The survey allows us to fill the gaps that neither log telemetry nor publication data can speak to, such as perceived time saved, bottlenecks, priorities, and changes in the scientific workflow. Together, the three data sources let us follow AI from capability, to adoption, to its effect on how science gets done.

Looking at the LLM interaction data, we find extensive penetration of AI in science. Use of LLMs in science is overrepresented relative to its share of US employment; for example, in the US, the Standard Occupational Classification (SOC) 19 job category roles (containing a set of Life, Physical and Social Sciences occupations) are about 2.7 times more likely to use AI compared to employment baseline. While more quantitative fields such as computer science and engineering lead in adoption, other specialties such as political science, medicine, and agricultural sciences are also significant adopters of AI. Specialized models are present across most disciplines, though the share varies substantially by field of study. They are also highly cited. Nearly half of the papers that introduce specialized models are in the top 1\% of citations within their respective fields. Usage in science is not sparse, with almost half of surveyed scientists using AI as part of their daily workflow. Looking at geographic variation in adoption, a one percentage point increase in a country's share of the world's researchers is associated with a 0.9 percentage point increase in its share of science LLM interactions, and specialized model development is concentrated in the U.S., China, the UK, the EU, Korea, and Canada. Some lower-income countries cite these models heavily without developing them, which suggests that open models can diffuse widely even where the capacity to build them does not.

Importantly, LLMs and specialized models appear to act as economic complements, filling out and potentially enhancing the capabilities of the other. While at the top level of the task taxonomy the two types of models look alike---quantitative data analysis and modeling dominates both---the similarity disappears once tasks are disaggregated further. The correlation in usage across tasks falls monotonically between the most aggregated and most granular tasks, and at the finest level there is almost no overlap. LLMs are used broadly across tasks such as writing and troubleshooting research code, statistical analysis, literature review, and helping with drafting documents. Specialized models are relatively more common in health and life sciences and are used to predict disease outcomes, engineer molecular constructs, and run complex simulations. Therefore both model families seem to reinforce one another in the production of scientific knowledge, limiting the extent to which one technology can at this point substitute for the other. Survey data tells a similar story: scientists split usage in both LLMs and specialized models.

Scientists also report seeing real productivity gains from AI. Around three quarters of researchers report time savings, with the average amount clocking in at almost 7 hours saved a week. That time goes back into research---into more output. About 8 in 10 report higher lab output over the past three years and 89\% expect further increases. The gains extend beyond speed. About 68\% of scientists report more access to insights from other disciplines, consistent with hopes that AI may lower the burden of knowledge that pushes researchers into ever-narrower specialties \citep{jones2009burden}.

But bottlenecks remain, which can explain why localized productivity gains have not yet translated into an explosion of new discoveries and applications. More than 4 in 10 of surveyed scientists report that their primary constraint has moved downstream over the past two years, into lab execution, clinical validation, and field data collection. AI tools may have increased the number of viable theories and hypotheses but not necessarily the means to test them. That has led 41\% to report a growing backlog of untested hypotheses. Verification absorbs a large share of the dividend: 89\% of those who save time spend more than a tenth of it checking AI outputs, and 46\% spend more than a quarter. Most concerning for the long run, 49\% of scientists say AI pushes them toward safer, more incremental projects where benchmarks are established and results are reliable, against 28\% who say it lets them take on riskier questions. If AI mainly lowers the cost of incremental work, it could raise the volume of papers while not meaningfully advancing the scientific frontier. 

This paper represents an early snapshot into how AI is affecting science right now. The division of labor between LLMs and specialized models points to where this could go: an LLM orchestrator that plans experiments, calls specialized models such as AlphaFold or a materials model, and interprets the results. That system is still work in progress. Delivering it and realizing its promise requires solving the bottlenecks that scientists identify and directing investment  to physical experimentation and validation capacity, to verification tools, and to the data and benchmarks that would let AI take on riskier questions rather than safer ones. 

The remainder of this paper is structured as follows. Section~\ref{sec:data} details our data sources and measurement. Section~\ref{sec:llm_use} shows some early signals of LLM adoption in science using Google ATLAS data \citep{iscenko2026atlas} mapped to scientific workflows and fields via the MIT FutureTech Scientific Task Taxonomy \citep{emmens2026taxonomy}. Section~\ref{sec:specialized_models} analyzes the fields, downstream task relevance, and early signals of spillovers of the specialized scientific AI models. Section~\ref{sec:survey} presents survey findings on researchers' time allocation, perceived bottlenecks, and expected productivity impacts. Finally, Section~\ref{sec:discussion} outlines our limitations and discusses the results.

\section{Data and Measurement}\label{sec:data}

\subsection{Preliminaries \& Scope}\label{subsec:preliminaries}

To build the basis of a comprehensive evaluation of AI in science, we combine three complementary data sources and two taxonomies: (i) a large-scale corpus of Gemini interaction logs classified into scientific use cases, (ii) an extensive inventory of specialized AI for science models, (iii) a primary survey of researchers, (iv) a new taxonomy of scientific tasks developed by MIT FutureTech, and (v) publication and citation records from OpenAlex. 

In order to measure ``science'' we have to define it. Throughout this paper, we adopt an intentionally broad definition of science. First, we include academic, industry, non-profit and government roles. Our data spans many fields including computer science, electrical engineering, molecular biology, animal science and zoology, sociology, and psychology: all are applied, computational, or theoretical. Second, we are informed by internationally recognized definitions of science \citep{ukri2025definition, uksciencecouncil2026, oecd2015frascati} and define science as any pursuit within these roles likely to generate new knowledge and understanding through investigation, computation, experimentation, or theoretical work. This is in line with the UK Science Council definition of a scientist \citep{uksciencecouncil2026}: ``a scientist is someone who systematically gathers and uses research and evidence, to make hypotheses and test them, to gain and share understanding and knowledge.''

To process, structure, and categorize the large-scale text, our pipeline relies on OCTO. This was previously used in \citet{iscenko2026atlas}, and is described in more detail there. At a basic level, OCTO integrates vector embeddings, clustering algorithms, and frontier Gemini LLMs to map unstructured text into structured representations at scale. OCTO serves as the method that classifies both logs and publication (abstracts) from our specialized AI model inventory against both the tasks and scientific fields hierarchies in a reproducible manner.

\subsection{Data Sources \& Taxonomies}\label{subsec:data_sources}

\subsubsection{Gemini Interaction Logs (Google ATLAS)}\label{subsubsec:atlas_logs}

To quantify science engagement with Gemini, we draw on the Google ATLAS 1.0 corpus \citep{iscenko2026atlas}: approximately 15 million anonymized interactions across the Google Gemini App, Google AI Mode, and API interfaces, sampled in early April 2026.\footnote{This sample does not currently contain paid use of Gemini API, which includes enterprise usage via Google Cloud. This limits our insights into LLM interactions in firms that signed under enterprise terms. This refers to specifically contracts signed under enterprise terms, and not e.g. accounts under institutional email addresses where no such contract was signed.}

We isolate likely scientific use cases through a three-stage filtering pipeline: (i) we use the ATLAS initial work versus non-work classifier to discard all non-work classified interactions (e.g. personal, leisure, and educational/homework inquiries); (ii) we use the ATLAS predictions of the likely occupations, applying a filter that restricts the sample frame to six 2-digit Standard Occupational Classification (SOC) categories spanning physical, life, social, computer, engineering, and health disciplines where research activities concentrate; and (iii) within this sample, we create a custom ``science'' classifier operationalizing the above definition of science. This step isolates, based on the content and context of the interactions between the user and the AI responses that are likely to be part of a scientist workflow.\footnote{Nonetheless, we acknowledge a limitation of this design which is that we do not know the identity of the user so it is not possible to know if a task is performed by a scientist or not unless the context hints at that, leading to potentially different biases in recovering different tasks (e.g. data analysis with data from an experiment will be easier to recover as science than some routine tasks helping debug equipment the scientist works with).} The resulting corpus yields about 360,000 science Gemini interactions. Using the OCTO classification framework \citep{iscenko2026atlas}, we map this corpus across two complementary taxonomies: the OpenAlex disciplinary hierarchy and the MIT FutureTech Scientific Task Taxonomy \citep{emmens2026taxonomy} across three hierarchical levels. Full sampling, procedure, validation and inherent limitations and uncertainties given the limited information for these predictions are provided in Appendix~\ref{app:atlas}.

\subsubsection{Specialized AI for Science Model Inventory}\label{subsubsec:model_inventory}

Alongside general-purpose LLMs, many scientists are relying on specialized AI models, such as deep learning and generative models for specific purposes like predicting protein folding. To track the development, diffusion, and task exposure of these tools, we constructed a curated inventory of 2,690 notable AI models for science.\footnote{The full inventory contains 5,501 models but we focus our analysis on models linked to an OpenAlex publication, published since 2012 and linked to an official code repository.} The inventory was assembled through agentic search across scientific repositories and literature, augmented by Epoch AI's Foundation Model database, and enriched with publication, citation, and institutional metadata from OpenAlex. Despite the efforts to make this comprehensive and unbiased in terms of domains covered, this is a work in progress dataset that should not at this point be treated as a ground truth dataset or exhaustive registry of specialized AI for science models.

To evaluate how these models could augment scientific research, we once again use the OCTO semantic classification tool to extract the downstream capabilities and tasks enabled by each model in the inventory (e.g., ``predicting protein structures'' or ``simulating fluid dynamics'') from its canonical publications' abstracts and map them directly to the MIT FutureTech Scientific Task Taxonomy and OpenAlex disciplinary hierarchy. Rather than capturing AI development activities, i.e. what the developers claim to be doing, our extraction focuses on downstream research tasks that can be ``touched'' (e.g. enabled, accelerated, or automated) by their model outputs. Full details regarding inventory construction, de-duplication, task-extraction and mapping heuristics, and validation are provided in Appendix~\ref{app:inventory}.

\subsubsection{Scientist Survey}\label{subsubsec:survey}

Since LLM logs and specialist AI model development and citations do not directly capture researchers' time budgets, potential bottlenecks in using AI, or perceived impacts (qualitatively, and quantitatively) we complement the data with a survey of more than 600 active scientists. The survey was run by a third-party provider through the More in Common online panel between 27 July and 11 August 2026 and aimed at researchers in the US and the UK. Participants were identified as scientists according to four criteria (detailed in the Appendix), which included that their main job involves working directly in science and technology, clinical or health research, life sciences, or social science research, and that they identify as someone who satisfies the aforementioned definition of science. 

These scientists span all of the primary scientific domains, including physical sciences, health sciences, life sciences, and social sciences and work in both academia/non-profits and industry. The survey collects detailed information on demographics, fields of research, and AI (incl. LLM and specialized models) adoption intensity. The survey also elicits self-reported time allocation across research tasks, expected and forecasted time savings from AI and perceived qualitative impacts on interdisciplinarity, tackling new questions, or bottlenecks. We note that due to the uncertainty and difficulty around capturing ``scientific'' usage, these survey results should not be interpreted as representative of the universe of scientists or to have a field or demographic breakdown in accordance with this universe. We present more details in Appendix~\ref{app:survey}.

\subsubsection{MIT FutureTech Scientific Task Taxonomy}\label{subsubsec:mit_taxonomy}

Our fourth data source is the MIT FutureTech Scientific Task Taxonomy, built from approximately 3.8 million \citet{lightcast2026job} scientific research job adverts posted after 2010. In their work, \citet{emmens2026taxonomy} define a scientific research job as one whose primary responsibility is to generate, or support the generation of, new scientific knowledge, therefore closely matching the definition applied when classifying Gemini interaction logs. The adverts are also classified into the OpenAlex subfield and field taxonomy covering academia, government, industry and non-profit research.

To build the taxonomy, \citet{emmens2026taxonomy} develop a pipeline that first extracts over 64 million task instances from the corpus of scientific job adverts. Often the same task is expressed in many near-identical ways across adverts, and therefore semantically equivalent task instances are consolidated into approximately 210,000 representative tasks. These representative tasks are organized under a global taxonomy of 12 Level-1 areas, 114 Level-2 areas, and 2,433 Level-3 groups. The hierarchy is built by combining hierarchical clustering of high-dimensional task embeddings, and LLM analysis. This taxonomy covers a range of responsibilities from broad areas such as ``Analyze and model quantitative research data'' to responsibilities outside of the lab such as ``Manage research projects and operational processes''. The taxonomy then maps from intermediate areas such as ``Analyze high-throughput experimental data'', to specific groups such as ``Analyze PCR amplification and assay data''. More details about this taxonomy and its limitations and validation can be found in \citet{emmens2026taxonomy}, as well as in Section~\ref{sec:discussion}. 

We also note that the fields in the MIT FutureTech Scientific Task Taxonomy are derived  from the OpenAlex (see below) list of scientific fields. For the purposes of this paper, we only focus our analysis on 217 scientific subfields that have research activities across both commercial (industry R\&D) and non-commercial (academia, government, non-profit) sectors, covering 92\% of all OpenAlex works, and 96.5\% of citations. Compared to OpenAlex, we exclude non-R\&D subfields and aggregate other subfields where the domain boundaries were not very sharp between fields, as also done in \citet{emmens2026taxonomy}. These exclusions and aggregations comprise mainly general umbrella classifications (e.g. Tourism and Hospitality, Industrial Relations) that also lack sufficient empirical R\&D job postings. 

\subsubsection{OpenAlex Publication Database}\label{subsubsec:openalex}

Finally, we use bibliometric records from OpenAlex \citep{priem2022openalex}. OpenAlex provides comprehensive metadata on scientific works, including publication dates, author affiliations, funding sources, citations, and concept classifications. We use these data to help create our specialized model inventory, as well as measure the diffusion of AI across disciplines. The MIT FutureTech Science Task Taxonomy is built on the subfield taxonomy developed in this database which facilitates comparisons across both sources. To address known issues in OpenAlex classifications (e.g., \citet{alperin2024analysissuitabilityopenalexbibliometric, Culbert2025}), we audit and clean metadata. For example, in section \ref{sec:specialized_models}, we use OCTO to reclassify papers (based on their content) into OpenAlex's subfields to address its tendency to disproportionately assign computational and data-driven papers under Computer Science. 

\section{Use of LLMs in Science}\label{sec:llm_use}

\subsection{Approach and Preliminary Observations}\label{subsec:llm_approach}

We start by analyzing the Google ATLAS 1.0 sample \citep{iscenko2026atlas}, containing 15 million anonymized user interaction logs across the conversational (Gemini App, and Google AI Mode) and Gemini API. We filter this focusing on work interactions for six major SOC categories where research activities concentrate (see sub-section~\ref{subsubsec:atlas_logs} and Appendix~\ref{app:atlas} for additional information). 

Our first descriptive finding is that the SOC codes that are likely to contain scientific interactions are significantly over-represented in the ATLAS sample compared to the general workforce. We measure this through an occupational over-representation ratio, calculated as an occupational group's U.S. share of total Gemini conversation volume divided by its baseline share of U.S. employment, calculated based on Occupational Employment and Wage Statistics (OEWS) data \citep{bls2025oews}. Across all six research-intensive occupational families combined, Gemini interaction volume was about 1.8 times higher than their share of the U.S. labor force. The concentration is most pronounced in the core scientific occupations (SOC 19-Life, Physical, and Social Science), where LLM usage is about 2.7 times higher than their employment baseline.\footnote{The core STEM detailed occupations within SOC 15, 17, and 19, had an over-representation ratio that was even higher, at 5.8x.} These numbers are even higher in core STEM fields.\footnote{These findings are in line with equivalent data from Anthropic's Economic Index \citep{appel2025anthropic}, where Life, Physical, and Social Science (SOC 19), as measured in June 2026 were also heavily over-represented (an over-representation ratio of about 4.5x in the US, and relatively close to the ATLAS v1.0 \citep{iscenko2026atlas} LLM interaction shares globally).} 

Since existing taxonomies are not geared towards identifying scientific work or workflows and we also do not know the identity or employer of our users, we implement on top of this filtering, the science classifier described in section~\ref{subsubsec:atlas_logs} with the goal of capturing a broad set of scientific or research activities (including academia, industry, government) which, based on the context and content of the interactions, are likely to represent scientific activities. 

From the initial sample of 15 million Gemini interactions, about 360,000 were classified as science LLM interactions. This is the sample we will focus on for the remainder of this section. We present more information about this sample in Appendix~\ref{app:atlas}. 

We find the proportion of science LLM interactions varies across occupational groups, ranging from 4.5\% in General Management (SOC 11) to 66.5\% in Life, Physical, and Social Science occupations (SOC 19). If we compare the science LLM interactions with the average work conversation, they are on average about +7\% more multimodal, about +19\% higher in token usage, about +11\% higher in number of turns, and measure +26\% higher domain expertise score (which measures in a categorical variable language sophistication and apparent domain knowledge; more details in \cite{iscenko2026atlas}).\footnote{This is based only on conversational surfaces (Gemini App and AI Mode) for which this data is universally available in order to be able to make these comparisons.} This suggests that these science interactions---perhaps in line with other frontier cognitive tasks---are more likely to be sophisticated compared to other Gemini work usage.

\subsection{Gemini Science Interactions Characteristics Across Fields}\label{subsec:llm_fields}

We now proceed to discuss variation across fields. We first map interaction clusters into the scientific disciplines of interest that are common and well-represented in the OpenAlex and MIT FutureTech Scientific Task Taxonomy using the OCTO classification algorithm. To carry out this work, we enrich the OpenAlex field descriptions with keywords and types of papers within each cluster, as well as contrastive language. For the MIT FutureTech Science Task Taxonomy we leverage the hierarchy and representative tasks by making use of analogous and distinct examples to delineate subfields.\footnote{We acknowledge some limitations of this approach in section~\ref{sec:discussion}, primarily that in the absence of knowing the user's identity prompts lacking specific context may be mapped to methodological fields (e.g., Computer Science or Mathematics) rather than the researcher's true field.}

\begin{figure}[htbp]
\centering
\includegraphics[width=\textwidth,height=0.72\textheight,keepaspectratio]{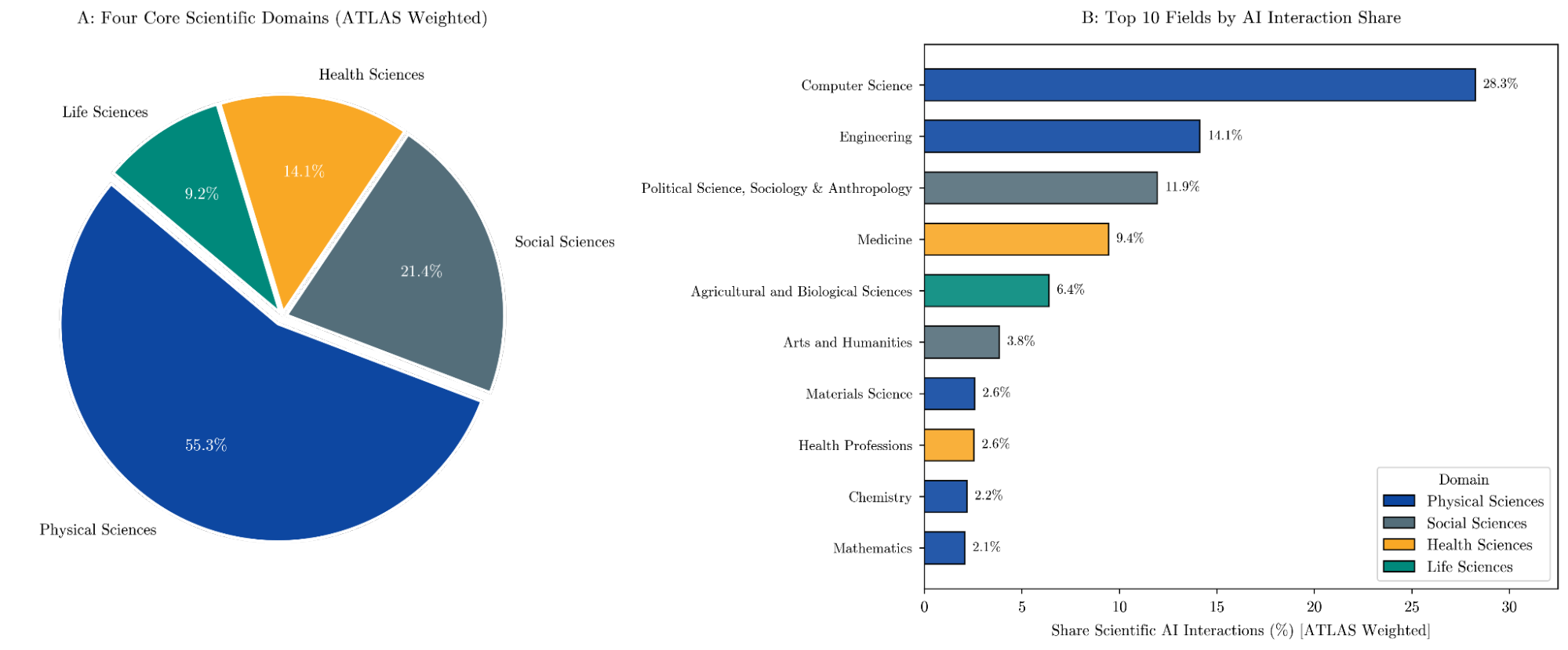}
\caption{Scientific Domain Representation in Gemini Interactions (including API)}
\label{fig:figure1}
\vspace{0.3em}
\begin{minipage}{\textwidth}
\footnotesize
\textbf{Notes:} This figure shows the distribution of science LLM interactions across scientific domains and fields. Panel A shows the disciplinary breakdown across four Level-1 domains (Physical Sciences, Social Sciences, Health Sciences, and Life Sciences). Panel B reports interaction shares for top 10 scientific fields. To distinguish the field from the overarching domain, the social sciences OpenAlex field (which includes political science, sociology, anthropology, education, public administration, etc.) is labeled Political Science, Sociology \& Anthropology.
\end{minipage}
\end{figure}

Using the \cite{iscenko2026atlas} strict privacy-preserving thresholds, we observe LLM usage in 195 of the 217 most granular scientific subfields.\footnote{The privacy-preserving multi-user threshold was set at 25 distinct users. This means that for the remaining 22 subfields, we can not distinguish if/what usage occurred.} This seems to suggest that AI adoption spans most scientific disciplines. Figure~\ref{fig:figure1} shows the breakdown by core scientific fields and domains. In Panel A, we show that science LLM interactions are led by Physical Sciences (including Computer Science), which account for more than half (55.3\%) of the scientific interactions. Social Sciences are the second-largest domain at 21.4\%, followed by Health Sciences, at 14.1\% and Life Sciences at 9.2\%. At the field level (Panel B), we show that Computer Science tasks lead all fields, accounting for nearly three out of every ten science LLM interactions, with Engineering (also a physical science), second at 14.1\%.\footnote{It is perhaps unsurprising that these fields come on top, as previous research tracing LLM-writing found both relative and absolute high adoption, particularly in Computer Science and related areas \citep{liang2025mapping,Liang2025, yu2026paperreviewedllmbenchmarking}.} Nonetheless, science-LLM interactions, as observed by our classifier, extend beyond the physical sciences. General Social Sciences (which includes Political Science, Sociology \& Anthropology) have around 12\% of interactions, ahead of Medicine, Agricultural \& Biological Sciences, and Arts \& Humanities. The remaining interactions span fields such as Materials Science, Chemistry, Mathematics, but also Economics \& Finance, Psychology, and Archaeology. At the bottom, we observe less usage in fields such as Dentistry and Veterinary Science.

This descriptive field breakdown does not take into account the number of potential researchers or how active these scientific fields are. One possible adjustment is to account for the number of unique researchers having at least one publication in OpenAlex in these fields. Here we acknowledge at least three limitations: a) this benchmark is only valid for scientists who publish their work, therefore underestimating the number of authors in fields with, for example, more industry employment and b) this benchmark does not account for differences in publication frequencies between fields, c) OpenAlex has several issues regarding mapping of scientific works (as will be discussed later in the scientometrics section). 

After normalizing usage by OpenAlex unique authors, we observe that a 1 percentage point increase in 2022 researcher share of a scientific field in OpenAlex data is correlated with an approximately 0.7pp increase in the share of the field in science-related Gemini interactions. Researcher shares explain just below 30\% of the variance in science Gemini interactions shares, reflecting that while the underlying number of researchers explains some field level variation, most AI usage is driven by other factors, likely including AI fit to particular tasks and researcher AI-openness. Physical Sciences and Social Sciences are generally over-represented in scientific Gemini interactions relative to their share of all science researchers\footnote{Computer Science sees more than 4x usage in science LLM interactions relative to its 2022 researcher count, leading field over-representation rankings.}, whereas Life Sciences and Health Sciences are under-represented.\footnote{Note that we only know the absolute numbers of requests, rather than having measures of interactions that are equivalent total labor hours. We also do not see task completion or saved time.} 

\begin{figure}[htbp]
\centering
\includegraphics[width=\textwidth,height=0.72\textheight,keepaspectratio]{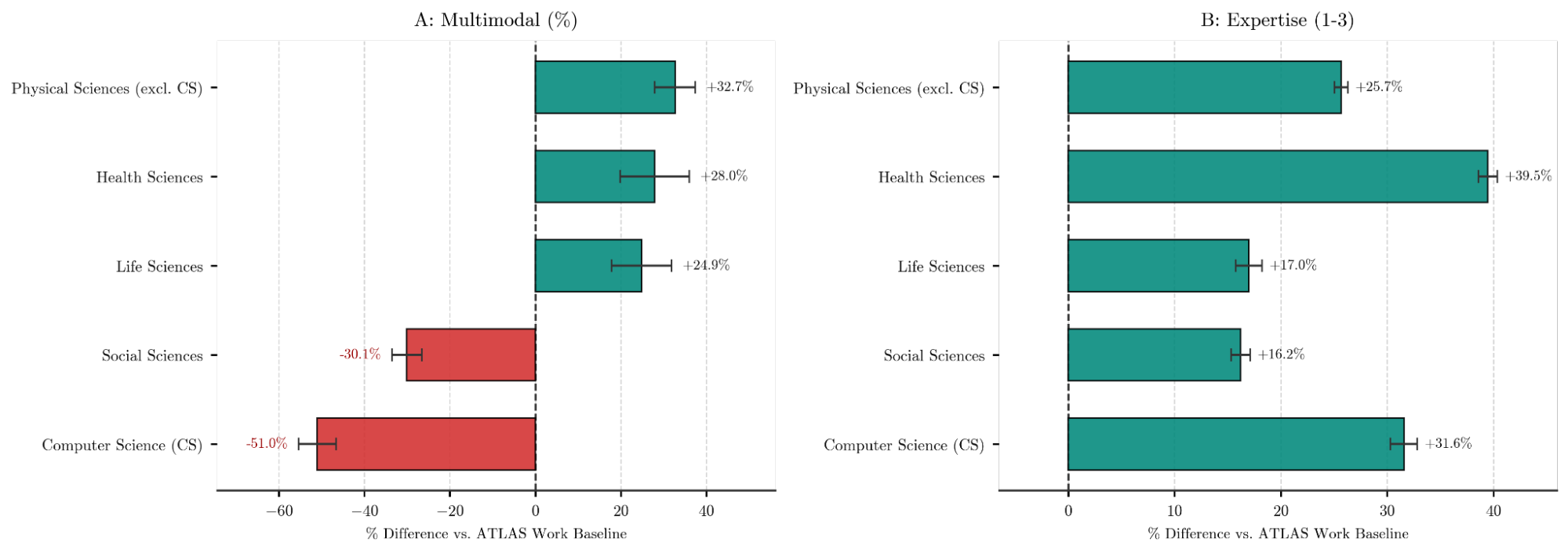}
\caption{Differences in Multimodal Use and Expertise by Gemini Interactions Scientific Domain}
\label{fig:figure2}
\vspace{0.3em}
\begin{minipage}{\textwidth}
\footnotesize
\textbf{Notes:} This figure calculates science AI conversation (Gemini App and AI Mode) characteristics against average work conversations. The two metrics are: (1) Multimodal Usage (\% of conversations retrieving or generating images/video); (2) Domain Expertise Score (categorical scale across all interactions calculated as in \citet{iscenko2026atlas}, a classifier that evaluates whether the query shows low, medium, or high domain knowledge). Computer Science is disaggregated from the remainder of Physical Sciences. Metrics are weighted using ATLAS sampling weights. Standard errors are computed via the delta method.
\end{minipage}
\end{figure}

Having now split the usage by field, we can look at heterogeneity along the intensive margin (i.e. unrelated to changes in the quantity of research) of usage as well (Figure~\ref{fig:figure2}). As expected, we find that physical sciences, life sciences, and health sciences tend to have higher multimodal average use than the average work interaction, while social sciences and computer sciences have comparatively lower shares (Panel A). Using the domain expertise score, as implemented in \citet{iscenko2026atlas},\footnote{This is not to be mistaken with the task expertise score developed based on \cite{NBERw33941}, which we plan to work with in the next iterations of this program.} we find that science Gemini interactions have higher expertise scores than average (Panel B) across all fields. Here, we take expertise to reflect user's prompt writing sophistication and domain impacts of AI in terminology fluency. 

\begin{figure}[htbp]
\centering
\includegraphics[width=\textwidth,height=0.72\textheight,keepaspectratio]{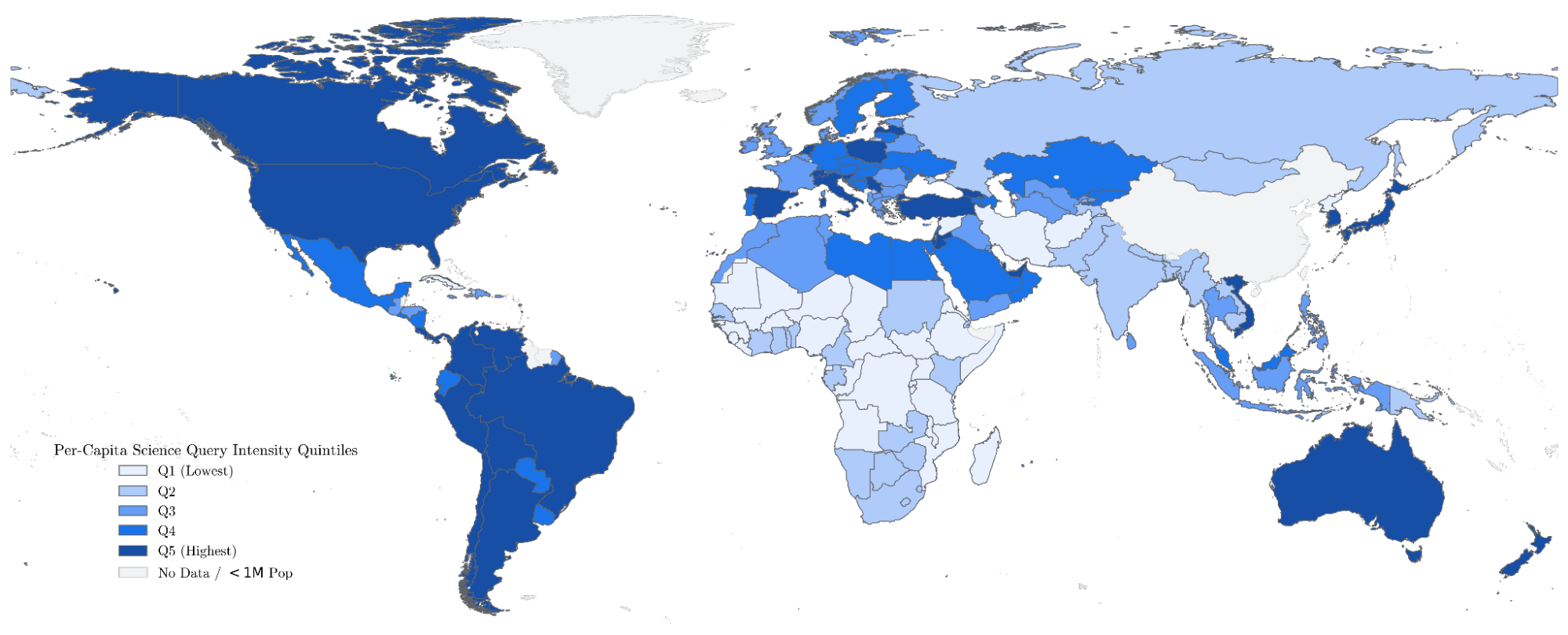}
\caption{Global Geography of Per-Capita Gemini Science Query Intensity}
\label{fig:figure3}
\vspace{0.3em}
\begin{minipage}{\textwidth}
\footnotesize
\textbf{Notes:} This map shows the global diffusion of Gemini science interactions across countries/regions with $\geq$ 1 million inhabitants and where data was available. China is excluded. Countries/regions are categorized into five intensity quintiles (Q1 Lowest to Q5 Highest). Data for population comes from the \citet{worldbank_population_2024}. 
\end{minipage}
\end{figure}

If we aggregate the number of unique OpenAlex authors in 2022 at the country/region level, we find that science LLM usage scales proportionally with the size of the science workforce. On average, a 1 percentage point increase in a country's/region's share of researchers is associated with an almost 0.9 percentage point increase in its share of science LLM interactions. Relatedly, in Figure~\ref{fig:figure3}, we show some per-capita heterogeneity. This simple regression explains about three quarters of the cross-country variance in science LLM interactions. Major hubs, like the US, South Korea, Japan, Australia, Singapore and Netherlands have high adoption. Conversely, countries/regions with smaller science workforce, including those in Africa, Central and South Asia, show lower science AI interactions.

\subsection{Tasks Breakdown}\label{subsec:llm_tasks}

Mapping LLM interactions to the MIT Scientific Task Taxonomy, we find that LLMs are used across the entire science workflow. ``Analyze and model quantitative research data'' has the largest individual share (about 42\% of science interactions), containing both code and theoretical work. This is especially high for Gemini used via the API. The second largest set of interactions, about 14\%, are under ``Communicating research findings and stakeholder information''. We note that we think that this is likely a lower bound, as such queries may have been more likely to be filtered out by the custom science classifier or the work-in-progress task mapping.\footnote{First, based on the data from the synthetic validation, our pipeline tends to over-predict writing tasks as ``analyze and model quantitative research data'' when these also include attachments or data mentioned. Second, routine writing (especially not related to work e.g. in personal emails), tone polishing, proofreading prompts may be more likely to be filtered out as generic text editing or misclassified in the wrong SOC, especially when they are short/without context. We plan to iterate on further improving this pipeline in the next iteration of this project.} 

Beyond communication and working with data, scientists also use Gemini as cognitive partners in physical tasks, teaching, and operations. Tasks in the ``Develop products, prototypes, and process technologies'' have 11.4\% of usage. ``Conduct experimental and sample processing operations'' have 9.8\% of usage. Scientists use AI for teaching and training (about 6.4\%), managing research projects (just below 4\%), and even to conceptualize research and manage and coordinate clinical studies. 

\begin{figure}[htbp]
\centering
\includegraphics[width=\textwidth,height=0.72\textheight,keepaspectratio]{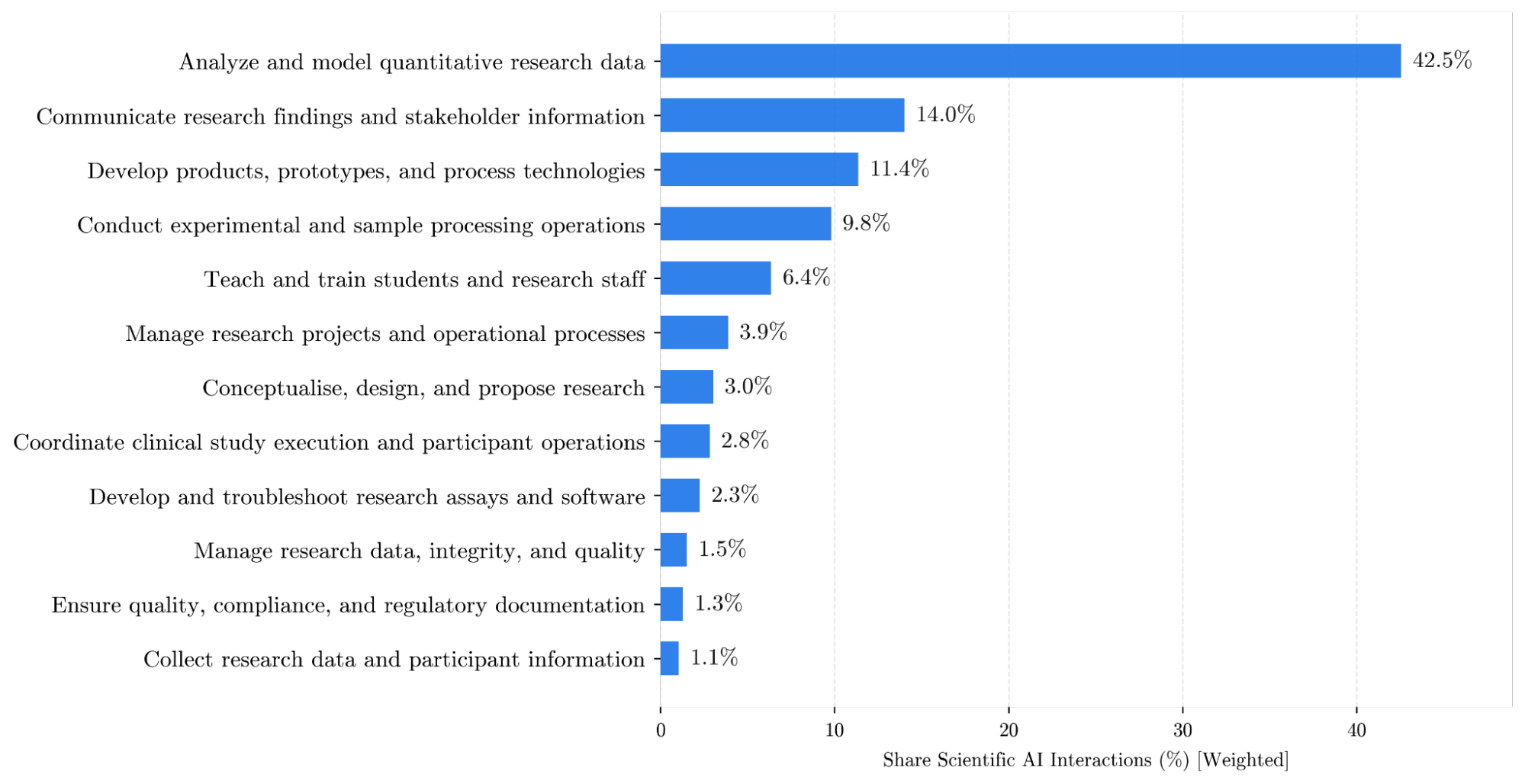}
\caption{Level-1 Task Representation in Gemini Interactions (including API)}
\label{fig:figure4}
\vspace{0.3em}
\begin{minipage}{\textwidth}
\footnotesize
\textbf{Notes:} This figure shows the breakdown of scientific tasks performed with Gemini, based on the scientific interactions sample as described above. Each scientific conversation cluster is classified into one of 12 Level-1 broad scientific task domains from the MIT FutureTech Scientific Task Taxonomy. Shares are weighted by conversation cluster weights (reflecting ATLAS interaction volume) and normalized to sum to 100\%.
\end{minipage}
\end{figure}

Lastly, we find that science LLM usage is heterogeneous across disciplines (some evidence in Table \ref{tab:table1}). Interactions about clinical studies are nearly 5 times more likely to appear as a share of activity in Health Science relative to baseline. Ensuring quality, compliance and regulatory documentation, as well as research proposal design are also significantly over-represented task types. In Life Sciences, experimental and sample processing tasks are over 3 times as prevalent. Social scientists use Gemini disproportionately for collecting research data (often using multimodal capabilities), teaching and training, as well as project management. Computer Science has a higher share of usage in troubleshooting software and for quantitative data analysis (coding), whereas Physical Sciences (excluding Computer Science) has relatively more usage for prototype and process technologies, as well as data analysis. We note that these metrics do not adjust for baseline workload shares across domains, so these figures may reflect workflow differences rather than differential task-level adoption propensities. Nonetheless, this provides some additional insight into how scientists use Gemini.

\begin{table}[htbp]
\centering
\caption{Most Relative  Over- and Under- Represented Tasks by Scientific Domain/Field}
\label{tab:table1}
\small
\begin{tabularx}{\textwidth}{>{\bfseries}p{0.22\textwidth} X X}
\toprule
\textbf{Scientific Domain} & \textbf{Over-Represented Tasks} & \textbf{Under-Represented Tasks} \\
\midrule
Computer Science (CS) & Develop and troubleshoot research assays and software; Analyze and model quantitative research data & Conduct experimental and sample processing operations; Coordinate clinical study execution and participant operations \\
\addlinespace
Physical Sciences (excl. CS) & Develop products, prototypes, and process technologies; Ensure quality, compliance, and regulatory documentation & Collect research data and participant information; Coordinate clinical study execution and participant operations \\
\addlinespace
Social Sciences & Collect research data and participant information; Teach and train students and research staff & Conduct experimental and sample processing operations; Develop products, prototypes, and process technologies \\
\addlinespace
Health Sciences & Coordinate clinical study execution and participant operations; Ensure quality, compliance, and regulatory documentation & Develop products, prototypes, and process technologies; Develop and troubleshoot research assays and software \\
\addlinespace
Life Sciences & Conduct experimental and sample processing operations; Manage research projects and operational processes & Collect research data and participant information; Coordinate clinical study execution and participant operations \\
\bottomrule
\end{tabularx}
\vspace{0.3em}
\begin{minipage}{\textwidth}
\footnotesize
\textbf{Notes:} This table shows tasks where AI adoption is relatively higher (most over-represented) or lower (most under-represented) compared to average researcher activity across disciplines. Tasks are defined using the 12 Tier-1 task areas from the MIT FutureTech Scientific Task Taxonomy \citep{emmens2026taxonomy}. Relative adoption intensity is calculated as the domain task share divided by the baseline average across the five domains (with Computer Science disaggregated from Physical Sciences), using ATLAS sampling weights. For each domain, the top 2 most over-represented and bottom 2 most under-represented tasks are reported.
\end{minipage}
\end{table}

\section{Specialized AI Models in Science}\label{sec:specialized_models}

\subsection{Model Inventory and Field Representation}\label{subsec:specialized_inventory}

We also analyze specialized AI model use in science. To do this, we have compiled and analyzed a set of 2,690 unique models that were published after 2012 and linked to a code repository suggesting that they could be used by the scientific community.\footnote{Depending on the model this might include details on model implementation and training, access to model weights and code to reproduce the publication where the model was introduced. We note that the size of this dataset is comparable to \citet{trisovic2025}, also based on Epoch AI data.} This inventory, which is similar to work on foundational models by \citet{trisovic2025}, was built through an agentic web search over public code repositories and scientometric databases, and augmented with the Epoch AI Models dataset \citep{epochai2026models}. We specifically include a broad set of both domain-specific AI models (e.g. protein structure predictions, genomic models, neural climate models, etc.) and deep learning models (e.g. neural networks, diffusion models, vision transformers) that could help scientific discovery, while excluding general statistical ML (e.g., regularized regression, random forests, etc.) and general software/ML frameworks (e.g., PyTorch, TensorFlow). See Appendix~\ref{app:inventory} for additional details. 

\begin{figure}[htbp]
\centering
\includegraphics[width=\textwidth,height=0.72\textheight,keepaspectratio]{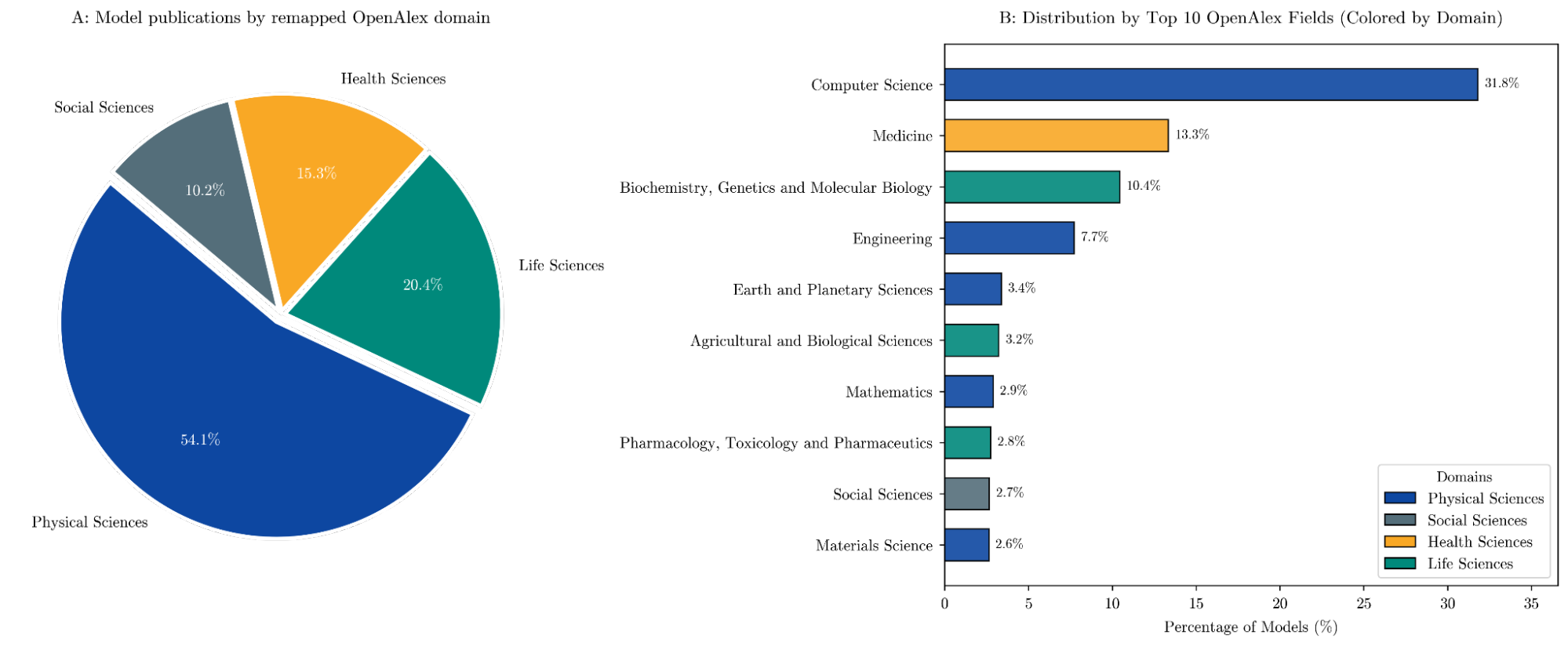}
\caption{Scientific Domain and Field Representation, Specialized Model Inventory}
\label{fig:figure5}
\vspace{0.3em}
\begin{minipage}{\textwidth}
\footnotesize
\textbf{Notes:} Panel A shows the share of models in the specialized inventory by OpenAlex domain focusing on models published since 2012 and an official code repository link. OpenAlex domains are based on an OCTO analysis of abstracts to avoid biases towards computer science and AI labeling in the OpenAlex data. Panel B shows the share of models in the specialized inventory by OpenAlex field focusing on the top 10 fields by overall level of activity. Each bar is colored by the domain where it sits. More information in Appendix~\ref{app:inventory}.
\end{minipage}
\end{figure}

The models in the inventory cover all 26 scientific fields in the OpenAlex taxonomy and 83\% of the subfields. Figure~\ref{fig:figure5} shows the percentage of models in each OpenAlex domain (panel A) and field, focusing on the top 10 fields (panel B). Although Physical Sciences / Computer Science dominate the inventory (as was the case in LLM use), Life and Health Sciences domains and fields within them account for more activity here -- almost 36\% of all models vs. 23\% in the LLM use data. Just above 10\% of the specialized models belong to Biochemistry, Genetics and Molecular Biology, a field that was not present in the top 10 fields in the LLM use data, and where it has been argued that AI has particularly large potential \citep{Topol2025, kohli2026understanding}.

\subsection{Specialized Model Tasks}\label{subsec:specialized_tasks}

We used OCTO to extract 4,475 tasks that specialized models help scientists perform based on their publication abstracts. These tasks are formatted using the same structure as the MIT FutureTech Scientific Task Taxonomy (verb / object / context). In total, specialized models are associated with over 200 unique ``actions'', highlighting the diversity of areas where AI can contribute to the research process. 

\begin{figure}[htbp]
\centering
\includegraphics[width=\textwidth,height=0.72\textheight,keepaspectratio]{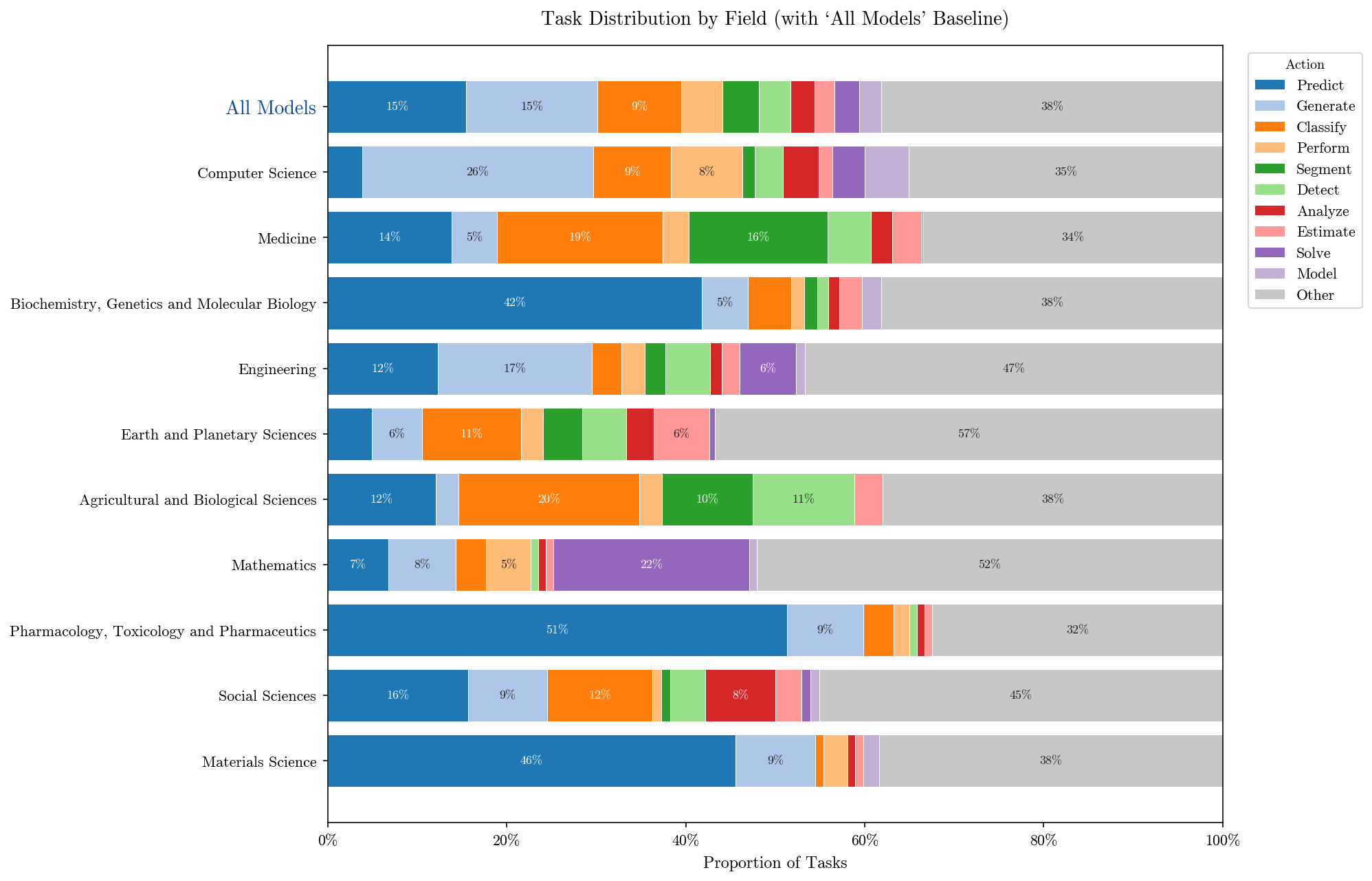}
\caption{Top Actions Performed by Models in Different Fields, Top 10 Actions and Fields}
\label{fig:figure6}
\vspace{0.3em}
\begin{minipage}{\textwidth}
\footnotesize
\textbf{Notes:} This stacked bar-chart shows the distribution of model task action verbs extracted from publication abstracts over scientific fields focusing on the top 10 fields by overall activity in the specialized model inventory, and top 10 actions. Actions in other categories have been relabeled as ``Other.'' Overall distribution of tasks is presented in the top bar. Social Sciences here refers to the general field (including Political Science, Education, Sociology \& Anthropology, etc.)
\end{minipage}
\end{figure}

Figure~\ref{fig:figure6} displays the top 10 action verbs associated with all models (top bar) and those in the top 10 fields by activity. It shows that specialized models are primarily used for prediction, generation and classification. Field heterogeneity seems to reflect variation in the relevant use cases in different fields. For example, specialized models in Computer Science are often used to generate text or images and video, while models in Medicine provide segmentation and classification capabilities for imaging data. Solving (mathematical proofs and theorems) is more important in Mathematics, and Analysis in the Social Sciences.

We then map the scientific tasks extracted from model abstracts to the MIT FutureTech Scientific Task Taxonomy with the goal of understanding what tasks and fields could benefit or be augmented with specialized model outputs and capabilities. When we look at the taxonomy as a whole, we find that 17.6\% of its Level 3 tasks are covered by models in the inventory. When we exclude operational and teaching tasks unlikely to be mentioned in publication abstracts we are extracting model tasks from, the percentage rises to 24.3\%.\footnote{We define here, in this context, operational tasks to include MIT Level 1 task domains such as ``Manage research projects and operational processes'', ``Teach and train students and research staff'', and ``Coordinate clinical study execution and participant operations.''} This means that almost 1 in 4 substantive research tasks in the taxonomy are covered by AI capabilities in our inventory.\footnote{Because specialized model capabilities are extracted from publication abstracts, they reflect the capabilities claimed by model authors rather than actual usage.} Of course this measure of ``coverage'' does not capture AI effectiveness in supporting a task. This would require analyzing model performance compared to established methods, an important avenue for future research.

\begin{figure}[htbp]
\centering
\includegraphics[width=\textwidth,height=0.72\textheight,keepaspectratio]{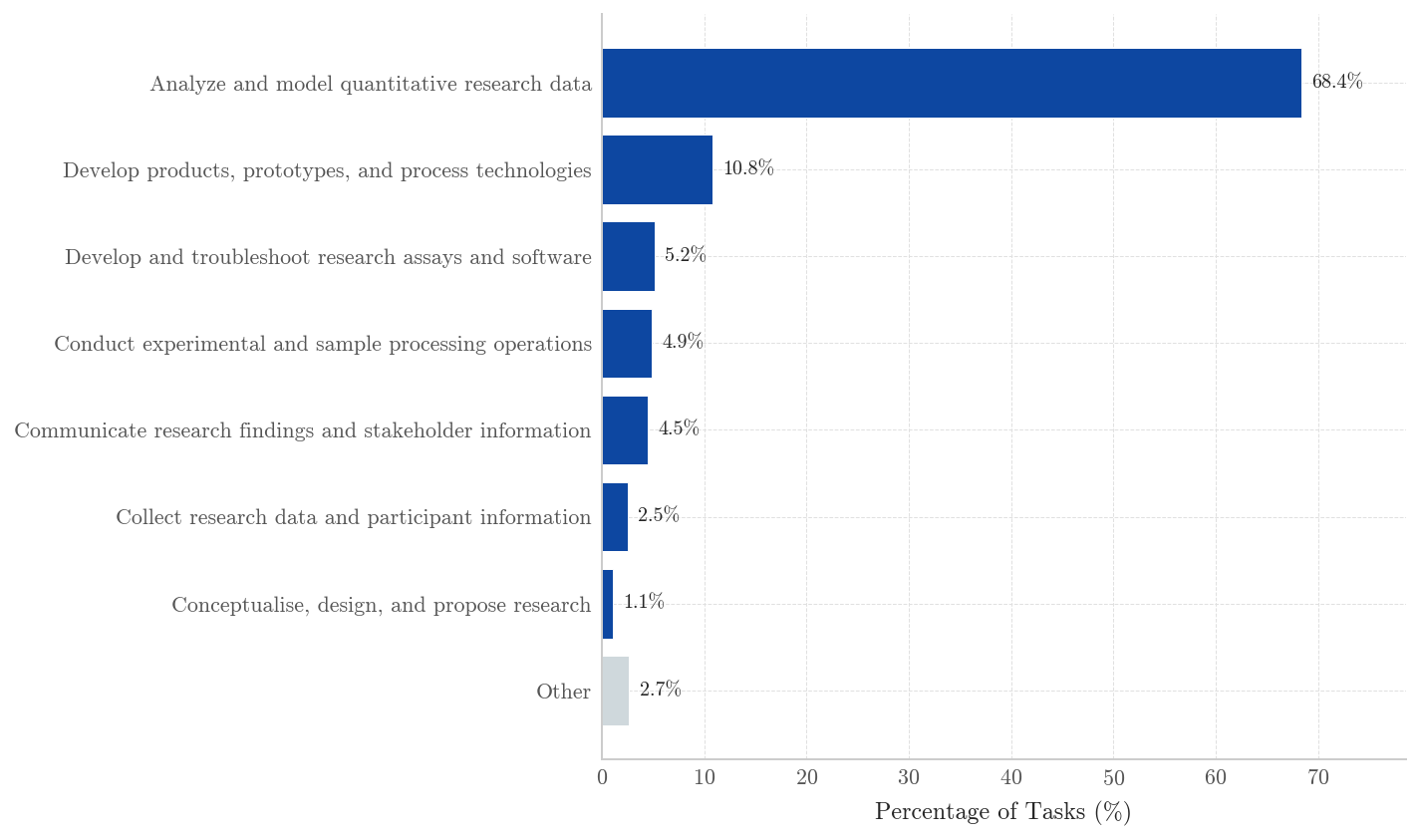}
\caption{Share of Specialized AI Models in MIT FutureTech Scientific Task Taxonomy Level 1}
\label{fig:figure7}
\vspace{0.3em}
\begin{minipage}{\textwidth}
\footnotesize
\textbf{Notes:} This bar-chart shows the share of specialized model tasks in MIT Level 1 domain categories. All domains with less than 1\% of activity in the data have been reclassified as ``Other.'' For more information, please check Section~\ref{sec:specialized_models} and Appendix~\ref{app:inventory}.
\end{minipage}
\end{figure}

Figure~\ref{fig:figure7} shows the percentage of extracted tasks mapped to taxonomy Level 1 Tasks. As it was the case with Gemini usage (Figure~\ref{fig:figure4}), ``Analyze and model quantitative data'' is the most common specialized model task. Looking at the more granular level into this category, one of the primary roles of specialized AI models appears to be the prediction and simulation of data (such as protein structures, genomic modalities and weather forecasts) that can be used in downstream analysis. In contrast to the Gemini log data, tasks related to research administration, operation and facilities or training are virtually absent from the specialized model inventory. We analyze this more formally in the next subsection.

\subsection{Division of Labor with LLMs}\label{subsec:division_of_labor}

When we consider the high-level task taxonomy, it looks like researchers tend to use specialized AI and general LLMs for similar task categories---the top usage category is data analysis and modeling for both. But if we zoom in on specific, detailed tasks, the picture changes completely. First, Gemini usage (as can be seen in Figure~\ref{fig:figure4}) includes tasks much more spread out across the entire scientific workflow: this includes writing, operations, and teaching. As we drill down into detailed tasks, the overlap between tasks covered by specialized models and Gemini becomes weaker. In practice, this means a scientist within the life sciences might use both specialized AIs and general LLMs under the broad umbrella of ``Data Analysis'' (Level-1). But at the more granular level (Level-3), some of the most over-represented fields in specialized models are ``Recombinant protein \& therapeutic engineering'', while for Gemini usage more common are tasks such as ``Statistical analysis \& hypothesis testing''. 

\begin{figure}[htbp]
\centering
\includegraphics[width=\textwidth,height=0.72\textheight,keepaspectratio]{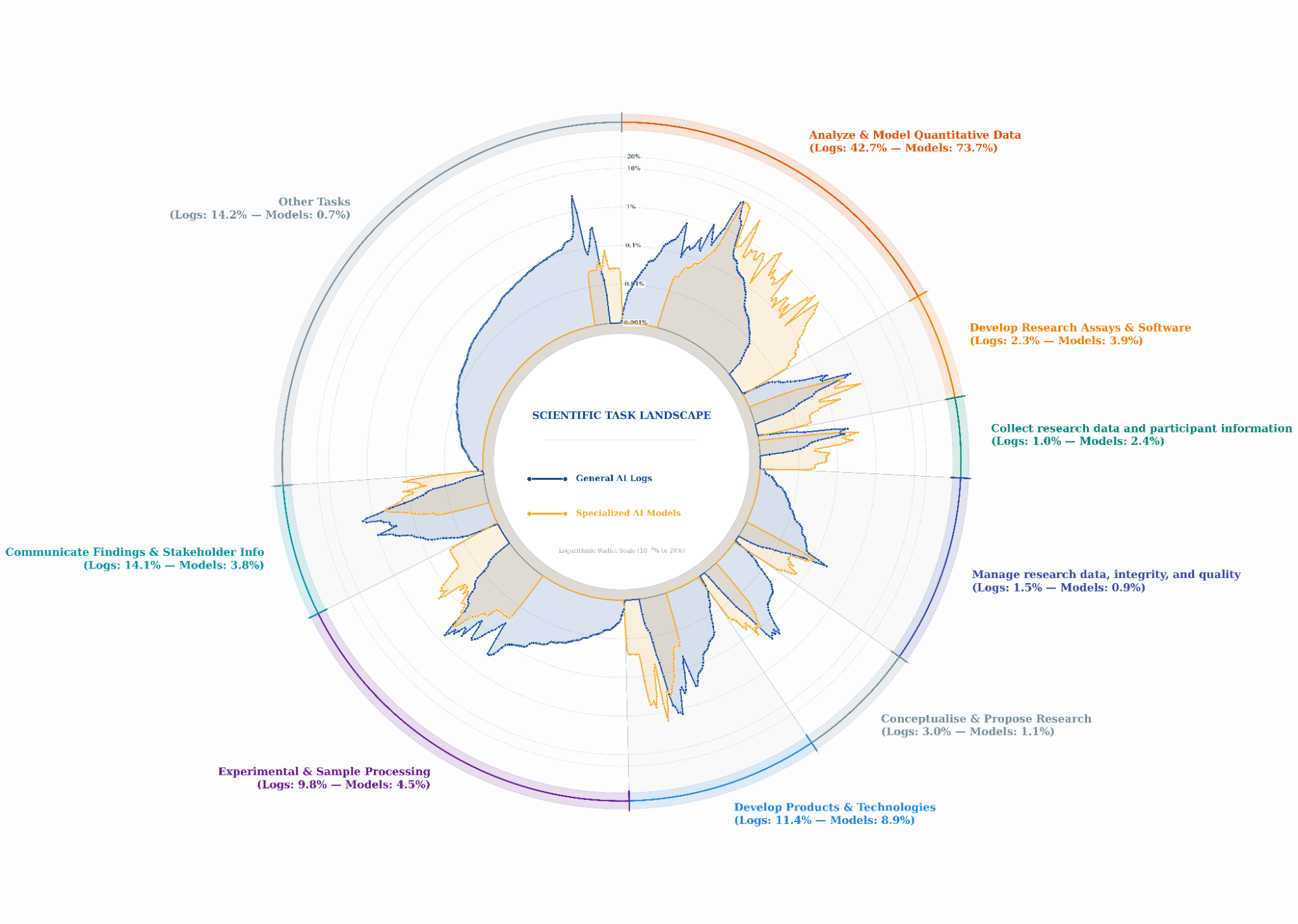}
\caption{Distribution of Science Tasks in LLM Interactions vs. Specialized AI Models}
\label{fig:figure8}
\vspace{0.3em}
\begin{minipage}{\textwidth}
\footnotesize
\textbf{Notes:} This radar plot compares the distribution of \citet{emmens2026taxonomy} Level 3 tasks in the Gemini usage (blue) and specialized models (yellow) datasets. Each point in the blue or yellow lines represents the share of all Gemini usage / specialized model tasks accounted for that Level 3 task. Tasks are arranged by the \citet{emmens2026taxonomy} level 1 domain they belong to. Level 1 domains are arranged by relative dominance of LLMs (left) or specialized models (right). We use a logarithmic scale and exclude \citet{emmens2026taxonomy} tasks that are not present in LLM logs or specialized models. We have combined all L3 tasks in L1 domains with less than 1\% of tasks in logs or specialized model inventory into an ``Other'' category. Note that when calculating the percentage of \citet{emmens2026taxonomy} Level 3 tasks in each Level 1 domain we remove Level 3 tasks classified as ``Other''---as these could be different qualitatively, which creates some divergences with the percentages reported in Figure~\ref{fig:figure7}, where we report all Level 1 tasks including their constituent Level 3 tasks classified as Other. 
\end{minipage}
\end{figure}

In Figure~\ref{fig:figure8} we compare the distribution of \citet{emmens2026taxonomy} level 3 tasks in the Gemini logs (blue) and specialized models (yellow) datasets. Points along each line indicate an individual Level 3 task's share of total activity within that corpus, plotted on a logarithmic scale. Tasks are clustered circumferentially by Level 1 domain, oriented with specialized-model-dominant domains to the right and LLM-dominant domains to the left. Tasks with zero activity in both datasets are omitted, and Level 1 domains with less than 1\% share in both sources are pooled into ``Other Tasks''. The non-overlapping contours of the radar plot show that specialized models and Gemini usage concentrate on different tasks of the scientific workflow. For example, specialized models (yellow) have a large spike within a concentrated cluster of quantitative data analysis tasks on the right. General LLMs (blue) form a perimeter across in different areas, which becomes more apparent in the ``other'' category that includes communication, teaching, and general operational support on the left.\footnote{As an example, some of the most over-represented tasks for specialized models are in disease and clinical outcome prediction, molecular construct and reagent engineering, and computational molecular modeling. Conversely, LLMs predominantly absorb a relatively higher share of simple hypothesis and statistical analyses and software troubleshooting and engineering. This is of course, in addition to the writing, communication, and workflow-support tasks like drafting manuscripts, and synthesizing literature, that are generally over-represented in all domains in Gemini logs compared to specialized models that don't generally help with such activities---as discussed in the main text.}

The log-log elasticity relating specialized model shares to general LLM usage shares drops monotonically as we increase the level of task resolution. The log-log elasticity is just above 0.6 at the most aggregate level, drops to 0.5 at Level-2, and is 0.2 at Level-3 when we only include tasks with positive shares across both specialized models and Gemini logs. If we include all tasks that show positive shares in at least one of the Gemini or specialized models, the elasticity remains 0.6 at the most aggregate level, drops to just below 0.5 at Level-2 and turns slightly negative at Level-3.\footnote{We caution that part of this decline is partially driven by attenuation bias: as granularity increases, classifier accuracy declines. While trying to adjust for this econometrically suggests that the elasticity remains lower even after accounting for attenuation, the nominal differences we present should be interpreted rather as upper bounds and in conjunction with the descriptive evidence in Figure~\ref{fig:figure8}.} This suggests that rather than being substitutes, general LLMs and specialized AI models may operate more as complements: researchers can deploy general LLMs as multimodal tools, coding assistants, and synthesis tools, while relying on dedicated domain models for the complex domain-specific hypotheses, data generation and simulations.

\subsection{Citations to specialized models}\label{subsec:citations}

The models in our inventory are highly cited, with more than 1.3 million total citations accrued since 2012, and almost half a million since 2020. This result is not driven by ``super-star'' models or different citation norms across disciplines---49\% of the models in our inventory are in the top 1\% normalized citations in their field, and 83\% in the top 10\%.\footnote{We note, however, that high citation counts could be partly built into our sampling frame: both our agentic search for notable tools and the Epoch AI database have either explicitly or implicitly a high impact or notability criteria in their sample. Rather than providing a definite quantification, we view these figures as qualitative evidence (or, even, a lower bound if there are more such models) of the significant reach and diffusion of specialized models across science.}

We have collected 460,000 unique citations to these models since 2020 and explored knowledge flows from disciplines that develop AI models to disciplines that cite models (an imperfect proxy for use); through this, we want to understand whether models and their allied outputs and techniques are diffusing into their ``home'' domains and / or flowing into other areas of science.

\begin{figure}[htbp]
\centering
\includegraphics[width=\textwidth,height=0.72\textheight,keepaspectratio]{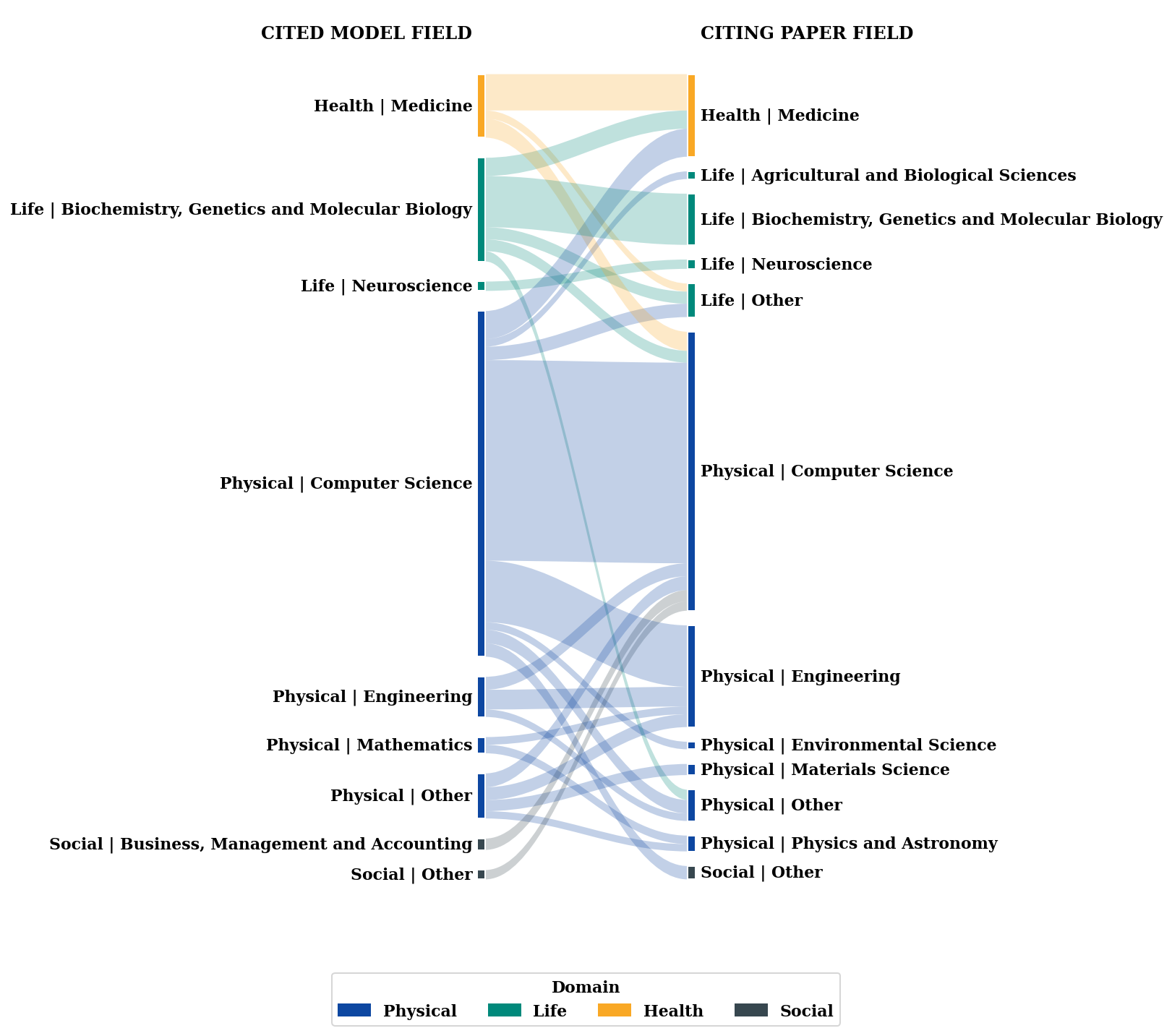}
\caption{Interdisciplinary Citation Flows}
\label{fig:figure9}
\vspace{0.3em}
\begin{minipage}{\textwidth}
\footnotesize
\textbf{Notes:} Citation flows from model inventory to citing papers recorded in OpenAlex. Papers were reclassified into scientific disciplines based on OCTO. We show these flows at the field level. To show the patterns of interdisciplinary knowledge transfer, fields with low citation counts are grouped into broader domain-level categories (e.g. ``Physical | Other'') and the sankey is further refined to exclude the flows with fewer than 5,000 citations. Computer Science is formally included in Physical Sciences. 
\end{minipage}
\end{figure}

We find that over a quarter of all citation links jump across scientific domain boundaries, suggesting a mix of within and between fields knowledge flows. Models developed in Health, Life and Social Sciences exhibit the strongest cross-domain knowledge flow---for example, almost half of citations to Health Sciences models come from Physical Sciences (dominated by Computer Science) and 10\% from the Life Sciences. Models with applications in the Social Sciences have over seven in ten of their citations into the Physical Sciences. By contrast, Physical Sciences models are more insular, with 85\% of citations from other Physical Sciences research. In Figure~\ref{fig:figure9}, we show some of these interdisciplinary citations flows across the most dominant fields.

There are several possible interpretations for disciplinary crossover: One could be reabsorption of domain-specific techniques into computer science; as an example, the case of the U-Net architecture \citep{ronneberger2015unetconvolutionalnetworksbiomedical,cicek20163dunetlearningdense}, which was originally designed for medical lesion segmentation but has since become popular across volumetric computer vision. In other cases, we might be capturing diffusion of knowledge between basic and applied fields, e.g. Life Sciences models such as AlphaFold, and health applications \citep{hill2026artificial}. We might also be capturing use of domain-specific data to benchmark architectures and techniques developed in Computer Science. Understanding the nature of spillovers and complementarities between specialized AI model development in different domains is an important question for further research.

\subsection{Global geography of specialized model development and citations}\label{subsec:geography}

We conclude by considering the geography of specialized AI model development and citations, based on the country/region of the last author of specialized model publications, and research articles citing AI models.\footnote{The results are broadly similar when we assign papers to their lead author or to the majority country among a publication's authors.}

\begin{figure}[htbp]
\centering
\includegraphics[width=\textwidth,height=0.72\textheight,keepaspectratio]{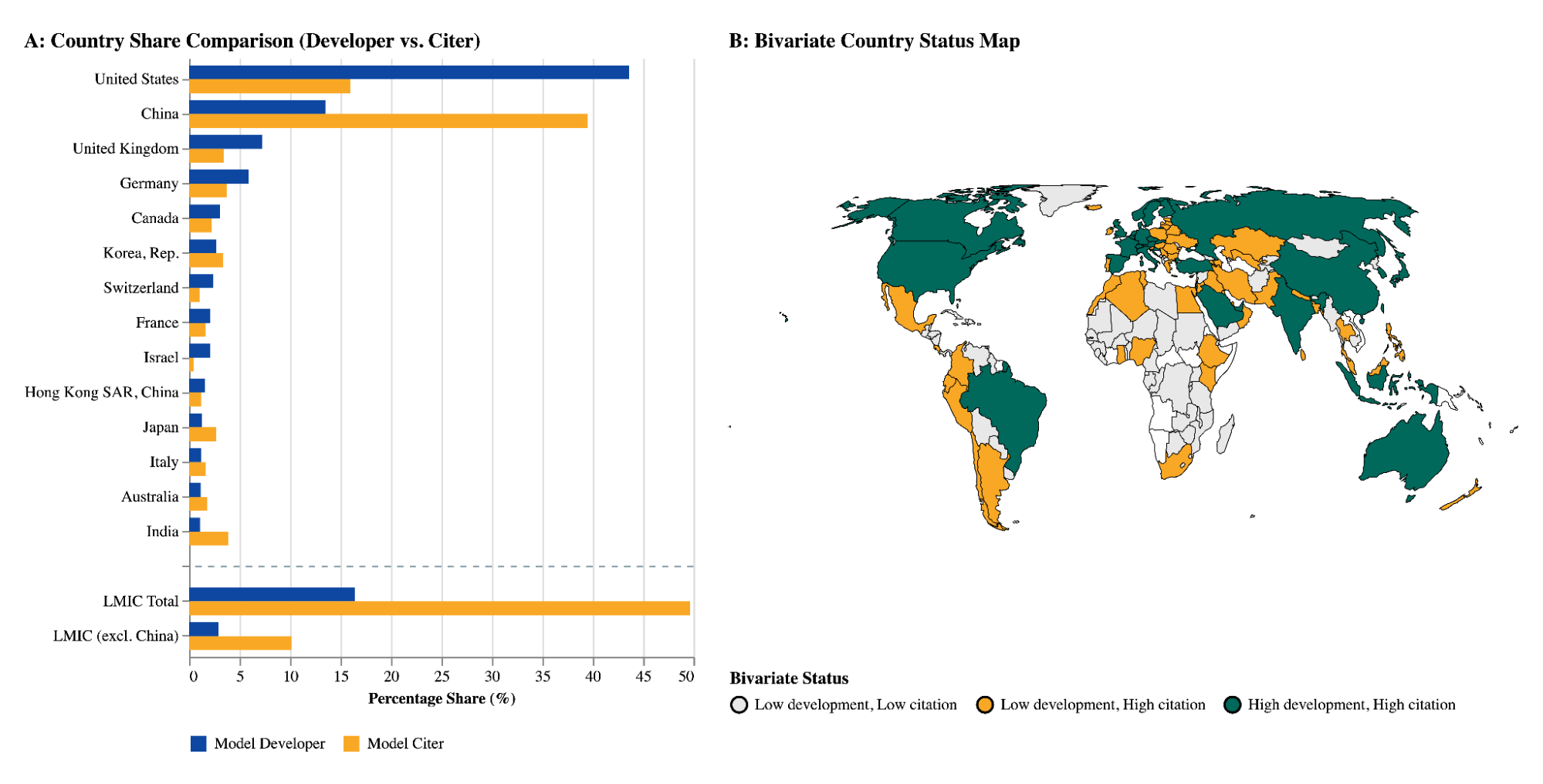}
\caption{Geography of Specialized Model Development and Citations}
\label{fig:figure10}
\vspace{0.3em}
\begin{minipage}{\textwidth}
\footnotesize
\textbf{Notes:} Panel A shows the share of publications linked to specialized models (blue), and citations to specialized models (yellow) by country/region or income group, focusing on the top 10 countries/regions in the specialized model inventory \textit{or} the specialized model citation dataset. Some countries/regions are present in both categories so the total number of countries/regions represented is less than 20. LMIC categories encompass the share of activity accounted for by Low and Middle Income Countries/regions according to the 2026 \cite{worldbank2026income} Definition including all and excluding China. Panel B shows a bivariate choropleth where each country/region is colored according to its joint position in the specialized model inventory and citation to specialized model distribution. Countries/regions above the median in both categories (e.g. US) are colored green, Countries / regions above the median in citations but not specialized model inventories are colored in yellow (e.g. Mexico). Countries/regions below the median (referred to as `low') in both are colored grey (e.g. Libya). Countries / regions in white are missing from the dataset.
\end{minipage}
\end{figure}

As Figure~\ref{fig:figure10} shows, there is a strong concentration of specialized model development in the US, China, UK, EU Countries, Korea and Canada. In fact, the top 10 model developing countries/regions account for 84\% of the models in our data - this is in line with the Gemini log analysis showing that countries/regions with larger scientific workforces participate in more scientific AI interactions. Low and Middle Income Countries (LMICs) are scarcely represented in the inventory with the exception of China, raising concerns that LMICs lack capabilities and resources to develop AI for science models tailored to their unique context, opportunities, and challenges. 

The picture changes when we consider the geography of countries/regions \textit{citing} AI models, which we treat as a rough proxy for specialized model usage. When we do that, we find that China accounts for around 40\% of citations of AI models and other LMICs such as India play a more active role in the data. This is also illustrated in the bivariate heatmap in Panel B of Figure~\ref{fig:figure10}, where a number of LMICs in Latin America, South-East Asia and Africa display a relatively low propensity to develop AI models but a higher propensity to cite them. One interpretation for this result worth exploring further is that while some LMICs might lack the resources to develop specialized models, they might still be able to use them in their research when these models are openly accessible.

\section{Survey Data on AI Impact on Science}\label{sec:survey}

\subsection{Survey Sample}\label{subsec:survey_sample}

Sections~\ref{sec:llm_use} and~\ref{sec:specialized_models} show two parts of AI adoption in science based on the task and field distribution of the usage in general LLMs and specialized AI models. Yet in order to assess the broader economic and organizational impact of AI, we need more information beyond usage patterns.

To gather these data, we partnered with an independent third party to collect a survey of 637 active scientists across the United States and the United Kingdom, spanning all the scientific domains previously discussed (Physical Sciences, Life Sciences, Health Sciences and Social Sciences) and across the whole career ladder (including Principal Investigators, professors, lab directors, industry R\&D managers, mid-level and junior research scientists). We used a four-level screening pipeline to isolate actual scientists (in accordance with the definitions in the other sections of the paper), but note that the sample may not be fully representative of the scientific population as a whole. Because global population-level administrative benchmarks across these diverse fields and ladders do not exist, we report unweighted results. More information can be found in Appendix~\ref{app:survey}.

\subsection{AI Usage and Extensive Margin Impact of AI}\label{subsec:survey_usage}

\begin{figure}[htbp]
\centering
\includegraphics[width=\textwidth,height=0.35\textheight,keepaspectratio]{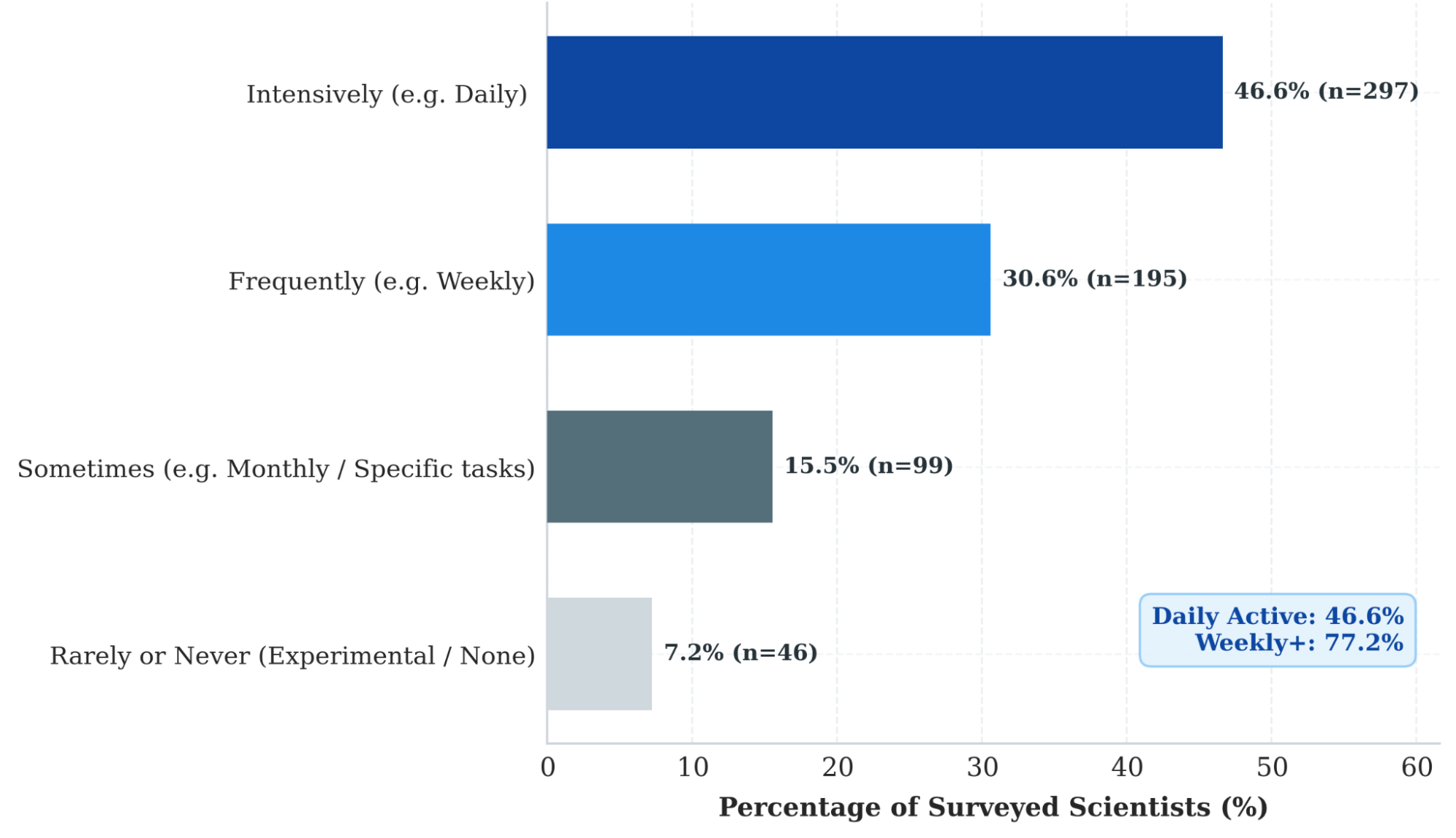}
\caption{Self-Reported Frequency of AI Use in the Sample of Scientists}
\label{fig:figure11}
\vspace{0.3em}
\begin{minipage}{\textwidth}
\footnotesize
\textbf{Notes:} Survey of 637 scientists in the UK and US conducted by More in Common in July-August 2026. The survey asked: ``How intensively do you currently use Artificial Intelligence (AI) tools---including Large Language Models (LLMs), specialized models, and coding assistants---in your scientific research?''. Response categories were: ``None at all'' and ``Rarely / Occasional experimental use'', which were merged in the category ``Rarely or Never'' for display purposes; ``Sometimes / Used regularly for specific tasks''; ``Frequently / A regular, integrated part of weekly workflow''; ``Very intensively / Core to daily workflow across multiple tasks'' and ``Almost exclusively / Research relies fundamentally on AI tools'' which were merged together in the ``Intensively'' category.
\end{minipage}
\end{figure}

We begin by examining if and how scientists interact with AI tools in their day-to-day work. In Figure~\ref{fig:figure11} we plot the frequency of AI use in our sample. We explicitly try to cover a broad definition of AI: including chatbots and document integrated LLMs, specialized AI tools and coding agents. We find that almost 47\% of the surveyed researchers use some form of AI daily, while another 31\% use it weekly.\footnote{While a lot of previous research primarily focused on LLMs \citep[e.g.,][]{naddaf2026nature, liang2025mapping}, our questionnaire also includes specialized AI models. While we prompted scientists to think of specialized AI models as tools or portals where they can access for example domain-specific AI, such as AlphaFold, GNoME, or DeepVariant \citep{Poplin2018}, we cannot rule out that some may have also included other areas not covered by Section~\ref{sec:specialized_models}, for example, traditional ML (e.g., random forests/clustering), computational software, or domain-tuned LLM wrappers (e.g. AI literature search engines) in their self-reported usage.} Surveyed researchers spend about 70\% of their time using general LLMs, split across General-Purpose Chat \& Document LLMs, which account for 41\% of total AI working time and the rest in Coding Agents and AI Code Assistants (previously studied in work such as \cite{lyttelton2026agents}). Specialized models account for around 30\% of self-reported time spent using AI, in our sample.

We also asked scientists how they are spending their work time in an average week. Looking at their responses highlights one possible reason why we may not yet see the propagation of task-level AI productivity improvements to more macro impacts on discoveries. According to our survey, in a typical working week scientists invest just under a fifth of their time on data analysis and interpretation (about 7 hours per week)---the set of tasks most associated with AI use (as discussed in Section~\ref{sec:llm_use} and~\ref{sec:specialized_models}). Surveyed scientists spend a large chunk of weekly time on data collection and experimentation (just below 8 hours per week), followed by writing and dissemination (about 5 hours per week), methodology and design (about 5 hours per week) and knowledge acquisition (about 5 hours per week). The rest of the working week is spent in conceptualization, as well as operational tasks such as funding \& administration (e.g. grants, management) that receive (together) about 9 hours in total. 

AI tools are self-reported to generate substantial net time savings in our sample. Following the established methodology of measuring firm productivity impacts of AI \citep{NBERw34836}, we find that just below three quarters of respondents report that AI saves them time on net in their working week, compared to only 6\% reporting net time lost. The average scientist's time savings are around 6.9 hours per week.

When asked how this saved time is reinvested, researchers report directing the dividend primarily into increasing overall research volume and output (just below 30\%), conducting physical lab execution and data collection (about 21\%), and tackling harder, more ambitious scientific problems (about 19\%). An additional about 18\% report improved work-life balance and fewer total working hours, while 12\% spend the saved time on teaching, mentoring, and admin tasks. 

\begin{figure}[htbp]
\centering
\includegraphics[width=\textwidth,height=0.35\textheight,keepaspectratio]{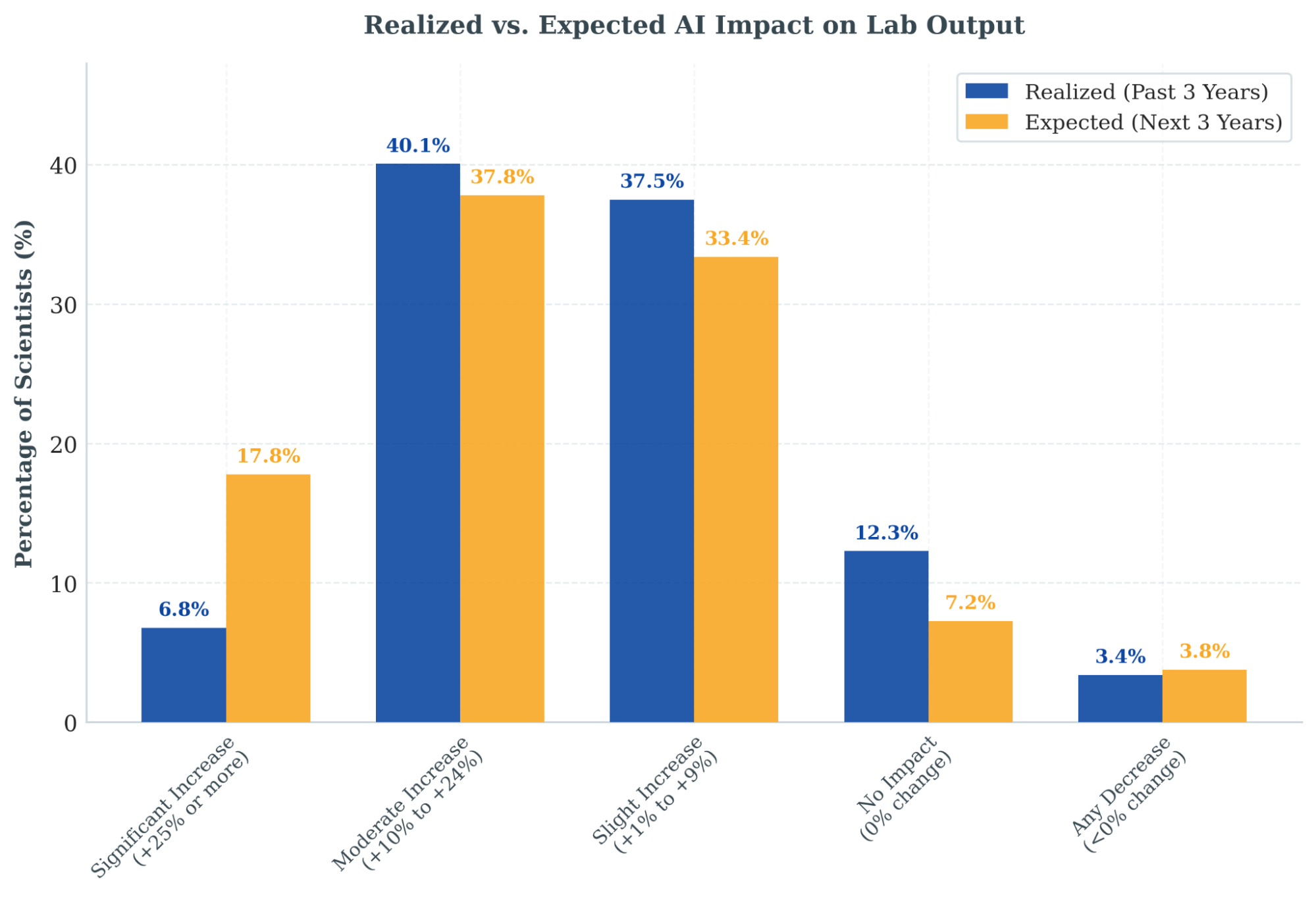}
\caption{Realized (past 3 years) and Expected AI (next 3 years) Impact on Lab Outputs}
\label{fig:figure12}
\vspace{0.3em}
\begin{minipage}{\textwidth}
\footnotesize
\textbf{Notes:} This figure compares the self-reported impact of AI on own laboratory and professional output over the past three years (realized, blue bars) and anticipated impact over the next three years (expected, orange bars) across surveyed scientists. The exact question we asked was ``How has the adoption of artificial intelligence technologies affected the number of professional outputs (papers, patents, discoveries, projects completed, etc.) in your lab over the past three years?''. Percentages reflect valid responses excluding `Other' (0.6\% in the realized question; 0.3\% in the forecast question).
\end{minipage}
\end{figure}

Consistent with evidence on the firm side \citep{BABINA2024103745}, over the past three years, about 84\% of surveyed scientists report a net increase in own lab or professional outputs (Figure~\ref{fig:figure12}), which includes more papers, patents, discoveries, and completed projects, with just below 47\% reporting a moderate or significant increase in outputs of 10\% or more. Looking forward over the next three years, about 9 in 10 scientists expect output increases, with the share anticipating the highest level of increases (25\% or more) moving from about 7\% to about 18\%. 

\subsection{Bottlenecks, Verification Costs and Scientific Research Question Impacts}\label{subsec:survey_bottlenecks}

\begin{figure}[htbp]
\centering
\includegraphics[width=\textwidth,height=0.72\textheight,keepaspectratio]{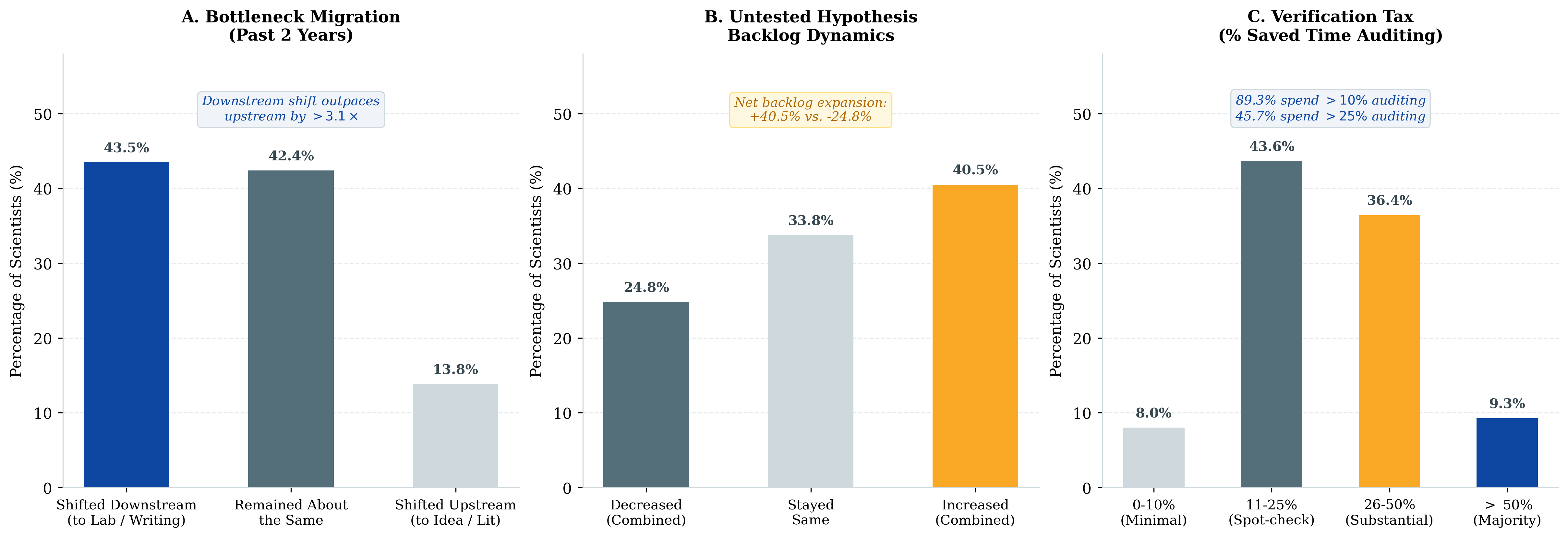}
\caption{Workflow Bottlenecks, Untested Hypotheses Backlog, Time Spent Auditing}
\label{fig:figure13}
\vspace{0.3em}
\begin{minipage}{\textwidth}
\footnotesize
\textbf{Notes:} Panel A shows the responses of question: ``Has this primary bottleneck shifted for you over the last two years?'' (asked immediately after surveyed scientists have identified their current rate-limiting bottleneck across seven research phases). Panel B shows responses to the question: ``Compared to three years ago, how has the size of your `untested hypothesis backlog' changed?'' We aggregated the increased (and decreased) moderately and significant categories in just one bar. Panel C shows the responses to the question: ``Thinking about the time you save using AI tools to generate text, code, or scientific hypotheses, approximately what percentage of that saved time is spent verifying, debugging, or fact-checking the outputs?''.
\end{minipage}
\end{figure}

As with other parts of the economy, the science production function can be characterized by bottlenecks at different stages of the research process. In our survey, we first asked researchers to identify their current primary rate-limiting bottleneck across seven research phases. Physical experimentation and data collection emerged as the largest bottleneck (24\% overall, and 30\% in Physical Sciences \& Engineering), followed by data analysis and interpretation (about 21\%), and writing and dissemination (about 14\%). We then asked whether this bottleneck had shifted over the past two years. Our survey provides empirical evidence of a downstream shift. In Figure~\ref{fig:figure13}A we see that about 44\% of respondents report that their primary bottleneck moved downstream to tasks such as physical wet-lab work or manuscript drafting, outnumbering those that say the past two years brought an upstream shift (about 14\%) by more than 3 to 1. For 42\% of respondents, the primary bottleneck remained in the same research phase. 

One interesting potential effect of this downstream shift is that about 41\% of scientists report that their backlog of untested hypotheses has increased, compared to about 25\% who reported a decrease (Figure~\ref{fig:figure13}B). One potential explanation is that AI accelerates hypothesis generation and computational predictions far faster than physical facilities can execute experiments to validate them, or the speed at which they themselves can verify the output. Indeed, a parallel change in workflow and new tasks is that AI outputs may impose substantial auditing and validation overhead (Figure~\ref{fig:figure13}C), with around 46\% of the surveyed scientists who saved time by using AI reporting that they spent more than 25\% of this saved time auditing and verifying AI outputs. Interestingly, this verification tax is particularly high in the Life Sciences. All these outcomes are positively correlated with AI use.\footnote{The downstream shift and untested hypothesis backlog growth are both statistically and economically very strongly correlated with AI use (in a series of linear probability models) on self-reported AI usage intensity, while the higher verification tax produces less robust, marginally statistically significant results, depending on specification.}

\begin{figure}[htbp]
\centering
\includegraphics[width=\textwidth,height=0.6\textheight,keepaspectratio]{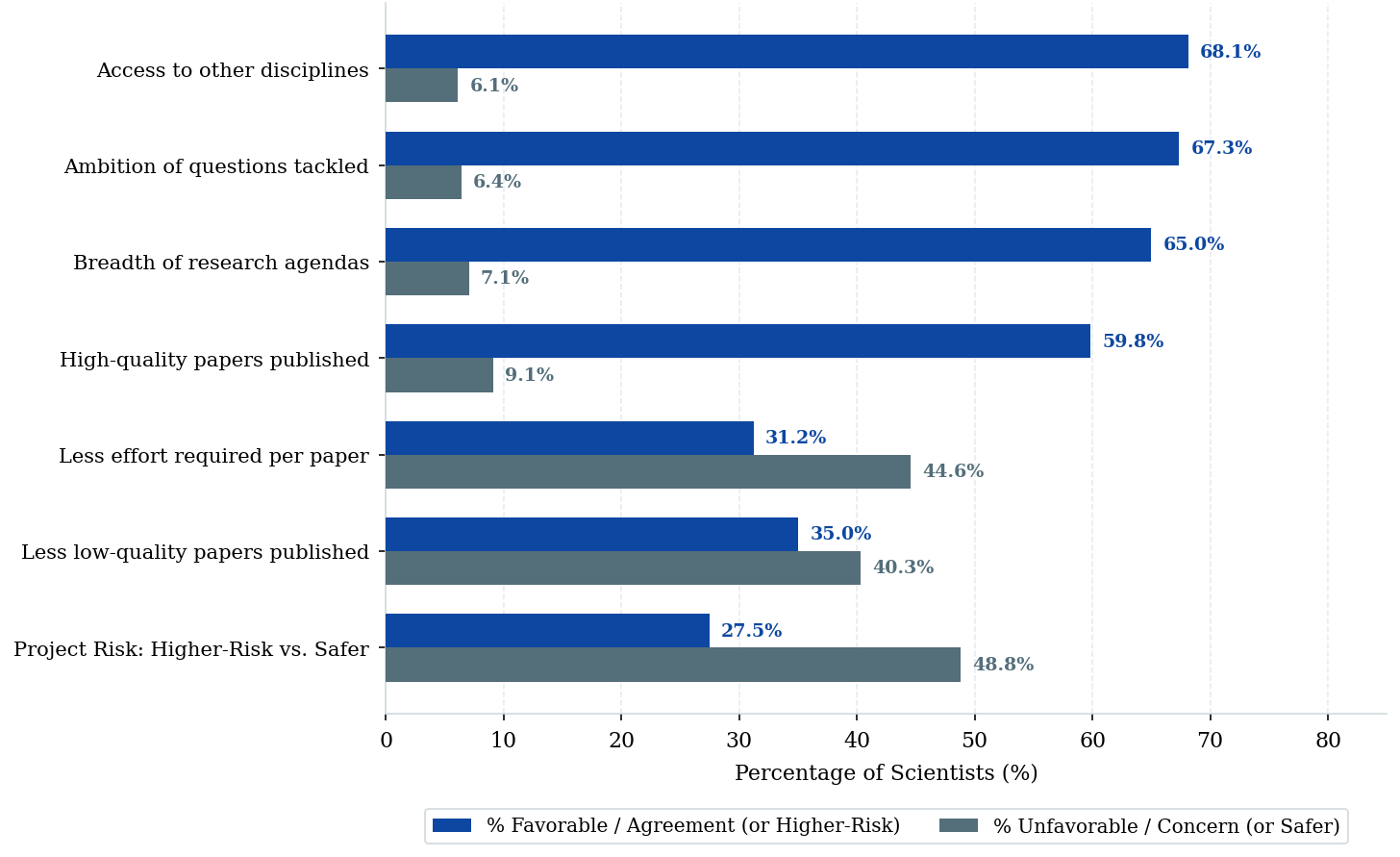}
\caption{Perceived Impact of AI on Science on Selected Outcomes}
\label{fig:figure14}
\vspace{0.3em}
\begin{minipage}{\textwidth}
\footnotesize
\textbf{Notes:} For the top six intensive margin impact questions, surveyed scientists were asked: ``Over the last two years, how has AI changed each of the following in your field, if at all?''. Respondents were given a five-point change scale. For the first four questions, blue bars show total increased (which is a combination of ``Increased a lot'' + ``Increased a little''), grey/darker slate bars show total decreased (``Decreased a little'' + ``Decreased a lot''), and neutral (``No real change'') which were excluded. For visualization purposes, for the fifth and sixth question (the effort required per paper and number of low-quality papers) we reversed the axis to show the favorable outcome (e.g. decreased effort) in blue and the unfavorable one (e.g. increased effort) in grey for consistency. For the last question the question was: ``How has your use of AI tools affected the risk profile of the research projects you choose to pursue?'' The blue bar shows the percentage saying ``It has encouraged me to pursue higher-risk, non-standard, or highly ambitious scientific questions'', while the grey/darker slate bar shows those saying ``It has encouraged me to focus on safer, more incremental questions where data and AI capabilities are well-established''. Neutral responses are excluded. 
\end{minipage}
\end{figure}

Finally, we asked scientists their perceived impact of AI on dimensions other than the quantity of scientific work. Their responses are displayed in Figure~\ref{fig:figure14}. Looking at scientists' perceptions of their field, about 68\% report AI increased access to insights from other disciplines, perhaps indicative of AI's ability to lower search costs across domains. Similarly, about 67\% report an increase in the ambition of questions tackled due to AI, and about 65\% report an increase in the breadth of their research agendas. At the same time, a significant number of surveyed scientists were concerned that AI is increasing the number of low quality papers (with 40\% saying that their number increased vs. 35\% who say these low-quality papers numbers decreased). Views on the effort to publish are also mixed: 45\% report increased effort per paper with some concerns for higher reviewer standards and validation requirements, while 31\% report decreased effort. Most concerning is that when evaluating their own work, a plurality of surveyed scientists said AI made them focus on incremental, safer questions, with 49\% seeing an increase in these safer questions vs. 28\% reporting an increase in higher risk work. Perhaps this is indicative of a ``Streetlight Effect'' \citep{nagaraj2023data, hoelzemann2024streetlight, hao2026artificial} where AI disproportionately lowers the cost of work in data-rich, tractable, benchmark-able problems while crowding out the exploration of the more unstructured, higher-risk questions where training data are scarce and (physical) validation is more costly.

Overall, this supports the idea that the impact of AI in science is not just about ``more'' output but also about ``different'' output: for example, we may expect to have both more interdisciplinary and ambitious outputs, but simultaneously we may also see lower quality or incremental papers as well. We plan to explore these questions further in follow-up work.

\section{Discussion and Next Steps}\label{sec:discussion}
\subsection{Limitations}\label{subsec:limitations}
Several limitations in our data, measurement, and scope must be acknowledged. First, our log analysis (section~\ref{sec:llm_use}) considered a sample of 15 million interactions drawn exclusively from Google's ATLAS \citep{iscenko2026atlas} dataset. Enterprise data was excluded from this investigation. Generalization should be done with caution as Gemini users may skew toward specific scientific disciplines or have distinct use patterns from other proprietary and open models. For example, scientists in more regulated areas like health sciences may be relatively more likely to use enterprise accounts or specialized solutions, which we cannot observe. It is also worth noting that our analysis excludes agentic AI tools like Antigravity where scientists can use Gemini and other LLMs to leverage specialized model capabilities \citep[e.g.,][]{applebaum2026science}---incorporating agentic data into our analysis is an important next step for our research program. 

Second, our specialized model inventory is an initial step towards capturing the wealth of domain-specific AI models in science, but at this point should not be considered representative. For example, we currently focus on notable models that are easier to detect through web searches, bibliometric platforms and coding repositories, but miss bespoke models designed during individual studies building on open source frameworks or fine-tuning open weight models, as well as commercial models. The specialized model inventory and the citation analysis building on it are also inherently lagging which means we are capturing a snapshot in a dynamic landscape. It could be that long-term citation patterns evolve differently as peer-review standards for AI-assisted papers change.

Our third data source---the scientist survey---could suffer from selection bias. In particular those scientists who are already enthusiastic about or frequently use AI might be more likely to respond to a survey about AI in science. This could potentially overestimate the real adoption rate and the average weekly time savings. Our sample size also limits our ability to explore disciplinary differences in key questions such as shifting bottlenecks in downstream validation (e.g., physical experimentation and clinical trials), which could vary by discipline.\footnote{While chemistry and medicine could face large capital/temporal barriers in physical validation, disciplines like computer science or mathematics can validate hypotheses entirely in-silico (computationally). Our aggregate findings, as a result, may over-represent the physical bottlenecks of ``wet'' sciences compared to ``dry'' computational sciences.}

We rely on MIT FutureTech's Scientific Task Taxonomy to integrate our analysis of LLM interaction log and specialized model inventory data. Deriving a task taxonomy from job postings has limitations. Defining the line between science research and the broader STEM economy is challenging, and there will be some false negatives and positives in the science research job postings used (and hence extracted tasks). The taxonomy will continue to be improved as we fine-tune the ability to characterize the tasks performed across scientific subfields. 

The mapping procedure we use to classify logs and tasks extracted from specialized models into the MIT FutureTech Scientific Task Taxonomy also has limitations. Our classifier currently has the inherent issues associated with trying to disambiguate what tasks are science (and all downstream field/task classifications) using LLMs, without knowing user identity. For example, we are explicitly trying to assign a ``science'' label (through the LLM making this classification) only when the content of the conversation seems to suggest it. As a result we may systematically omit implicit or loosely framed scientific work differentially across task types (e.g. context may be less prevalent when scheduling meetings, and more prevalent when doing data analysis for a paper). 

Additionally, and leaving aside classification error emerging from the need to assign noisy descriptions of scientific activity ``in the wild'' to ambiguous and in some case overlapping labels in a large taxonomy, our classification procedure maps each log/model task to a single MIT task, ignoring the fact that, for example, some specialized tasks might be relevant for multiple downstream scientific activities, or multiple tasks might be performed in a single Gemini conversation. It might also be the case that scientists ``find their own use'' for open source models, leveraging them for tasks that the developers had not anticipated or mentioned in the model descriptions. We plan to address these limitations and improve our classification and filtering procedure in future work.\footnote{Our reliance on abstracts to extract specialized model tasks is also likely to neglect a long tail of tasks involved in the design and implementation of scientific methodologies. Our decision to focus our specialized model analysis on the AI model capabilities being produced rather than the R\&D activities to build those capabilities will under-represent back-end AI and ML development tasks that are becoming more important as AI is adopted across science. We will give these tasks more attention in future work.}

We also note our paper captures a snapshot of AI adoption through the early and mid-period of 2026. This may matter for some of our findings. For example, we found a clear division of labor between LLMs and specialized models in this period. However, this may be subject to change. The emergence of new frontier LLMs (and specialized models) increasingly able to solve difficult problems in fields like mathematics, genomics or life sciences \citep[e.g.,][]{anthropic2026claude, callaway2026nature, openai2026rosalind, romeraparedes2024mathematical} creates uncertainty about the future evolution in the division of labor between different types of model. We plan to track this in future iterations of our work.

Finally, our results establish associations rather than causality. Future work is needed to isolate the causal impact of using AI tools on scientific productivity.

\subsection{Discussion}\label{subsec:discussion}

Notwithstanding these limitations, we believe that our findings provide an important early look into how AI is reshaping science. Rather than a simple story of labor substitution or runaway productivity gains from automated discovery, our evidence points to a more nuanced picture.

AI usage is widespread across scientific disciplines, so it is perhaps unsurprising that scientists appear more over-represented in terms of AI usage relative to other occupations. Of course, the diffusion of AI in science is closely related to other distributions (and capacity to make investments, e.g. \citet{10.1257/mac.20180386}) in computational or intangible capital. Gemini usage scales roughly in proportion to national researcher populations and the creation of specialized AI models is mostly concentrated in a small group of developed economies. Cutting-edge scientific AI often requires complements like compute availability and domain specific human capital. This suggests a risk that without policy action under-resourced or developing regions could fall behind. The frontier of AI-augmented science might require investment to mitigate global disparities in research capacity. 

In terms of task analyses, specialized AI models and Gemini usage appear similar. Both have the primary use category in quantitative modeling and data analysis. But when we dig into specific types of analysis, we start seeing evidence of model specialization and a division of labor within the scientific production function. Specialized models function as a kind of domain-specific capital---generating synthetic data, predicting molecular properties, and executing simulations---tasks that seem better served by a specialized model instead of a general LLM (at least for now). Meanwhile, general-purpose models absorb a wider range of tasks like writing code, synthesizing literature and operations. Similar to the way in which researchers specialize across disciplines or workflow components, AI tools may be specializing as well.

Where adopted, AI seems to deliver meaningful productivity dividends. Researchers report substantial net time savings that they recycle back into other projects. However, these changes enabled by AI in science will not only be on the extensive margin. About half of surveyed researchers report that AI directs them toward safer, incremental questions where clean data and benchmarks exist. Only 28 percent instead pursue riskier projects--despite evidence that most important scientific progress relies on high-risk exploration \citep{Azoulay2011} and novel hypotheses \citep{Uzzi2013}. It remains to be seen whether AI models will lead researchers toward more incremental work, enable ``moonshot'' projects, or possibly both; future work should examine the relationship between AI usage and the risk profile of scientists. It might be that AI delivers a ``Streetlight Effect'' \citep{hoelzemann2024streetlight, nagaraj2023data} where the costs of executing incremental work drops relative to higher-risk projects. Incremental work made easy by AI could grow as a proportion of research output overall, at least in the short-run.\footnote{This is not necessarily a negative outcome---for example, incremental ``normal science'' that enhances the understanding of disease mechanisms contributes to evidence on drug development \citep{mcnamee2017timelines}.} On the other hand, scientists report that AI enables more interdisciplinary work, potentially acting as a translator helping researchers recombine ideas more effectively \citep{fang2025generalization} and helping overcome the burden of knowledge \citep{jones2009burden}. 

Despite widespread task-level time savings, macro-level scientific discovery rates remain bound by workflow task dependencies. These hidden scientific ``organizational'' complementarities are similar to AI's impact on work in a general sense \citep{demirer2026chaining, NBERw34639, kremer1993oring}. For example, it may be that computational modeling and data analysis are accelerated by AI tooling or that specialized AI models provide more promising biological structures to test, but the residual tasks in the bundle of scientific work are harder to scale and absorb the time savings. This bundling and task interlock for scientific work means that broader productivity gains may require redrawing job boundaries, and for some types of work the task bundling is stronger than others \citep{garicano2026weak}. Enhanced productivity in data analysis might, for example, push the queue of work into hard-to-automate, physical stages that become a bottleneck \citep{jones2025artificial}.

Furthermore, AI outputs are not necessarily ``turn-key''. Almost 90 percent of researchers who report time savings spend a meaningful share of their saved time verifying AI outputs, and about 46 percent spend more than a quarter of their saved time doing so. As with software engineering, scientists who adopt AI tools may be pushed toward auditing, debugging, and quality screening. Especially in the case of scientific work, being correct is of paramount importance. Lowering the marginal cost of hypothesis generation, code generation, analysis, or testing will create a high complementary marginal value of verifying results (or, perhaps, human judgment e.g., \cite{AGRAWAL20191}). This high ``verification tax'' likely arises from the high value of reliable and correct output in science. 

The division of labor we document also suggests what the scientist of the near future might look like. Today a researcher moves between tools by hand: a chatbot for code and drafting, AlphaFold or a materials model for prediction, a spreadsheet or a lab notebook in between. The next step may be an LLM that orchestrates the specialized models directly, planning the experiment, calling the right model for each step, checking outputs against each other, and returning a candidate answer for the scientist to judge. The human's job shifts toward choosing the question, setting the standard of evidence, deciding what is worth the cost of a physical test. Our data say we are not there yet: verification still consumes a large share of the time AI saves and the backlog of untested hypotheses is growing. The hope, in the end, is larger than time savings. When AlphaGo \citep{Silver2016} played its 37th move against Lee Sedol in 2016, it played a move that its own model estimated a human would make about one time in ten thousand, and it won. The promise of more advanced AI systems in science is that they will eventually make moves of that kind: discoveries beyond the limit of what human researchers would have tried on their own.

Our work here has provided some initial measurements of the adoption of AI tools in science, its potential impact and bottlenecks that could limit this impact. As these limitations arise, so do new opportunities for cross-disciplinary work and new directions of inquiry in computationally demanding domains. Scientific production shares many of the features of other kinds of production. The rapid integration and adoption of AI tools by scientists combined with the substantive importance of breakthrough discoveries makes this an ideal domain to understand opportunities posed by AI. By continuing this work over time, we can watch the experimenters experiment, learning from scientific successes and failures to unlock AI-driven productivity gains in scientific R\&D and elsewhere in the economy.

\pagebreak 

\bibliography{references}
\newpage
\appendix
\renewcommand{\thesection}{\arabic{section}}
\titleformat{\section}{\normalfont\Large\bfseries}{Appendix \thesection:}{0.5em}{}

\section{LLM Sample Procedure and Validation}\label{app:atlas}

The LLM interactions analyzed in this paper are coming from the Google ATLAS 1.0 sample \citep{iscenko2026atlas}. We have approximately 15 million anonymized interaction logs across Gemini App, AI Mode, and API. Enterprise and paid customer API traffic are excluded from the sample frame.\footnote{We acknowledge that due to the exclusion of enterprise data, these likely reflect just one slice of scientific work, which is generally more sensitive towards privacy and data sensitivities and it is also possible that certain scientific domains (like health sciences) are more affected by these sensitivities.} The logs are subjected to a privacy-preserving pipeline that is fully explained in the original paper: prompts are stripped of personally identifiable information (PII), proprietary data or attachments, and identifiers. The data is subsequently aggregated into conversation clusters and then classified based on the content in work and non-work, point at which each goes through a more detailed SOC-O*NET or ATUS classification. 

Because the underlying ATLAS sample is predominantly composed of non-science related tasks, we isolate likely science LLM research interactions through a three-stage pipeline. In the first stage, we filter the work interactions (to separate them from non-work like educational interactions). On synthetic validation benchmarks based on \citet{iscenko2026atlas} designed to replicate a variety of user prompts, this classifier achieves a 93.7\% accuracy rate. 

In the second stage, to prevent contamination from unrelated occupations, we restrict our work sample to six 2-digit Standard Occupational Classification (SOC) major groups where science interactions may be most likely. This classifier from \citet{iscenko2026atlas} had an accuracy level based on synthetic data of 71.6\%. We include: Life, Physical, and Social Science (SOC 19), Computer and Mathematical (SOC 15), Architecture and Engineering (SOC 17), Healthcare Practitioners and Technical (SOC 29), Educational Instruction and Library (SOC 25), and Management (SOC 11) occupations. 

Within this subsample, we apply an LLM classifier operationalizing the standard definitions of research and R\&D as defined in the main text, section~\ref{sec:data}. To be a positive scientific match, a conversation cluster must show, in either the context or the underlying tasks as described in the privacy preserving cluster summary, that they are working toward the conception, creation, or discovery of new knowledge. Positive criteria include examples of the Level-1 tasks from MIT FutureTech Scientific Task Taxonomy. Conversely, negative criteria include context showing this as a routine execution lacking a novel research component/context. This is a conservative step intended to exclude activities like standard commercial IT and software engineering (e.g., basic web development), and operations that are not part of a scientific workflow (e.g., scheduling, accounting, writing reports, etc.) that are common in the base sample so could overwhelm this analysis if broadly included. Filtering through this pipeline isolates about 360,000 science interactions, with the proportion of confirmed LLM interactions ranging from the lowest of 4.5\% in general Management (SOC 11) to the highest of 66.5\% in Life, Physical, and Social Science (SOC 19).\footnote{We caution these reflect the specific sampling and decisions in the ATLAS sample \citep{iscenko2026atlas} and should not be thought of as an exhaustive census of scientific R\&D.} 

Each labeled science conversation cluster is processed through the OCTO for classification. At the discipline level, clusters are mapped into 4 canonical domains (Physical Sciences, Life Sciences, Health Sciences, Social Sciences)\footnote{We also added a formal ``Other Sciences'' option to minimize errors from trying to predict science from incomplete information.}, 26 primary fields, and 217 subfields. These come from OpenAlex. In parallel, clusters are also mapped into the MIT FutureTech Scientific Task Taxonomy \citep{emmens2026taxonomy} across Levels 1-3.

We then further validated the new measures (on top of those already validated in \cite{iscenko2026atlas}). We first evaluated the science classifier against a newly created set of science LLM interactions, produced using the same methods as \citet{iscenko2026atlas}. The classifier had 96\% accuracy. We also tested the classifier against labeled negative control conversation clusters, representing other non-science ATLAS conversations. We obtained an accuracy rate of 97\%. Finally, we looked at accuracy rates in the true science sample for tasks and fields. These were naturally lower, which is expected given the much higher size of the sample to match to: up to 2,433 for tasks at Level-3 level and 217 for subfields. These ranged from (conditional on achieving the right level) accuracy rates at each level of between 30\% (Level 3)-65\% (Level 1) for tasks and 75\% (fields)-83\% (domains) for disciplines. We aim to improve these rates in the future. Our checks confirmed that most discrepancies are in semantic overlaps in adjacent and interdisciplinary subfields and tasks, especially in settings with less context/information, and broad categories like analyzing and modeling data. One of the main weaknesses is that without information on the identity or employment of the user this classification is uncertain and clearly bounded by the amount of context provided in prompts.

\section{Specialized model inventory collection \& processing}\label{app:inventory}

We have built a specialized AI for science model inventory with the goal of mapping AI capabilities that could enable and augment scientific R\&D, focusing on tools that are openly available for scientists. The inventory includes deep learning, generative AI, foundation models and architectures but excludes statistical ML techniques such as regularized regression, random forests, clustering etc. that have also played an important role in scientific R\&D for decades. 

The inventory was assembled, expanded, and audited in three main phases. First, we used agentic search to find notable AI models across OpenAlex subfields. For each subfield, parallel researcher agents ran a discovery procedure querying GitHub topic tags and curated lists and searching open-access literature and web sources. This included validation checks to ensure extracted URLs were live and not hallucinated. 

Second, we used Epoch's AI models dataset, a regularly updated dataset with models that ``are notable for advancing the state of the art, or a large impact on the world or the history of the field.'' We collected a snapshot of this dataset in August 2026 and merged it with the agentic database after removing duplicates.

Third, we extracted and deduplicated DOIs and arXiv links from the expanded dataset and queried them against OpenAlex. We extract publication metadata for those including abstract, affiliation of first and senior (last) author and majority country/region in authorship list, citation metrics and primary OpenAlex topic, field and subfield. We also queried Semantic Scholar to fetch abstracts for papers with DOIs but no OpenAlex match.

Finally, as OpenAlex had substantial errors in the discipline mapping, we used OCTO to re-classify models into the OpenAlex domains, fields and subfields where their outputs might be used, reducing OpenAlex's tendency to e.g. assign models to computer science fields because of their use of computational and algorithmic methods. 

The final version of the inventory contains 5,501 models (4,858 after deduplicating models with different parameter sizes, variants etc.). Just under 70\% of these are linked to OpenAlex publications, and 57\% to an open source code repository (generally in GitHub) often containing model implementation and/or weights and reproductions. The inventory contains over 400 models developed by Google or Google DeepMind - over three-quarters of these are in Computer Science. For the analysis presented in this paper, we further restricted this sample to only models that were published after 2012, the year AlexNet \citep{alexnet} was published, and had an associated paper and official code repository, leaving us with a sample of 2,690 models.

The inventory does not seek to, at this point, provide an exhaustive or representative list of AI models for science. By definition, it focuses on notable and/or large models, which creates a bias towards visible, higher profile or better resourced research. It does not include general purpose deep learning frameworks such as PyTorch or TensorFlow that can be used to develop publication-specific models, and it will under-represent proprietary and close-sourced commercial models that might be used in industry and in some cases are not publicly revealed. It includes many AI models developed by AI researchers which might be less focused on enabling scientific R\&D in other disciplines. Although we were able to link almost 70\% of the models in the data to OpenAlex works, we note there are gaps in their OpenAlex metadata for preprints, particularly in variables such as normalized citations or author institutional affiliation.

We then use OCTO faceting capabilities to extract 4,475 tasks from the 2,690 unique model descriptions that remained after filtering, and then map the extracted tasks to the MIT FutureTech Scientific Task Taxonomy and to the 3-level OpenAlex domain - field - subfield taxonomy. 

During task extraction and mapping, we are most interested in identifying AI model capabilities and outputs that might be relevant for downstream research (e.g. AlphaFold 2 generates structural data for biologists and drug development) than on the tasks performed by the scientists who developed the models (e.g. AlphaFold 2's developers created innovative algorithms and trained their model on Protein Data Bank) which are also described in abstracts - so we adjusted faceting and mapping prompts accordingly. We also fine-tune the task extraction and mapping prompts used by OCTO to reduce its propensity to classify tasks into computer science and ML subfields instead of the application domains that might benefit from model outputs. 

One important limitation of our mapping procedure is that it assigns to each AI for science extracted task a single MIT task. This neglects the fact that AI capabilities could enable multiple downstream tasks (e.g. AlphaFold 2 can be used to interpret experimental data, develop biological hypotheses and simulate protein behavior c.f. \citet{kovalevskiy2024alphafold}). We hope to address this important limitation through one-to-many task mappings in follow-up work.

We validate our data extraction and mapping procedure by assessing its ability to recover the number and type of tasks and fields in 1,000 synthetic, scientifically plausible abstracts describing AI for science models with a known number of specific tasks in the MIT FutureTech taxonomy.\footnote{When seeding tasks for abstracts in a subfield, we only consider those that appear with high frequency in the subfield according to MIT's analysis of online job ads. This reduces the risk of generating implausible abstracts if we randomly matched tasks to subfields.} 

Focusing on task extraction, our pipeline extracts the exact number of tasks used to seed the synthetic abstracts in 48\% of the cases. Error is driven by over-extraction (47\% of cases) where our pipeline extracts more tasks than the number used to seed the abstract. We analyze over-extraction outcomes with Gemini and find that in almost all cases the tasks extracted were entailed by the abstract and not hallucinated. One explanation is that during the process of generating plausible synthetic abstracts, our method introduces additional tasks beyond the initial seed list. Focusing on mapping accuracy, we are able to recover the exact MIT task between 62\% at the broadest level and 28\% of cases at the most granular (over 7x to $>$600x improvement versus a random baseline), and the exact OpenAlex disciplines between 74\% and 48\% of instances (about 3.7x to about $>$100x improvement versus a random baseline). Although there is clearly room for improvement in our classification pipeline, our prototype pipeline achieves results comparable to ATLAS in a challenging multi-label classification task where errors are often driven by ambiguities and overlap between tasks and field definition. 

\section{Supplementary Survey Information}\label{app:survey}

More in Common interviewed 637 scientists in the US and UK on behalf of Google DeepMind, with fieldwork conducted online between 27 July and 11 August 2026. Respondents were recruited through specialist research panels and passed four screens: their main job had to involve working directly in science and technology, clinical or health research, life sciences or social science research; they had to actively conduct scientific research or R\&D as a primary part of that job; they had to confirm that their job fits the UK Science Council definition of a scientist, someone who systematically gathers and uses research and evidence to make hypotheses and test them, to gain and share understanding and knowledge; and anyone working in market research, marketing or advertising, or management consultancy was screened out. There is no reliable population frame for scientists in either the UK or the US, so we think of this as a screened non-probability sample rather than a representative one, and the data are therefore reported unweighted with no margin of error quoted. Survey findings reflect the self-reported, aggregated responses of independent researchers participating in their professional capacities and do not represent the official positions or institutional endorsements of Google, GDM, their respective employers, academic universities, or research institutes. 

Our total sample size was 637 active scientists across the United States (379 respondents) and the United Kingdom (258 respondents), split across the four canonical scientific domains quoted in our paper: Physical Sciences \& Engineering --including Computer Science (230 respondents), Life Sciences (167 respondents), Health \& Clinical Sciences (147 respondents), and Social Sciences (93 respondents). In terms of seniority and career stage, 356 respondents have senior leadership roles as Principal Investigators, Professors, Lab Directors, or Industry R\&D Managers (comprising 240 in industry and 116 outside industry), 234 respondents are mid-career researchers (128 industry and 106 outside in positions such as postdoctoral researchers and staff scientists), and 47 respondents are early-career (e.g. junior researchers and PhD candidates). Respondents report an overall sample average of just below 13 years of research experience.

\end{document}